\documentclass[aps,prd,preprint,superscriptaddress]{revtex4-1}
\usepackage{graphicx}
\usepackage{amsmath}
\usepackage{amssymb}
\usepackage{slashed}

\def\m0{m^{\!\!\!\!^o}}
\def\be{\begin{eqnarray}}
\def\ee{\end{eqnarray}}
\def\bc{\begin{center}}
\def\ec{\end{center}}

\def\Re{{\rm Re\,}}
\def\Im{{\rm Im\,}}

\begin{document}

\title{On coupled-channel dynamics \\in the presence of anomalous thresholds }


\author{M.F.M. Lutz}
\affiliation{GSI Helmholtzzentrum f\"ur Schwerionenforschung GmbH, \\Planckstra\ss e 1, 64291 Darmstadt, Germany}
\author{C.L. Korpa}
\affiliation{Department of Theoretical Physics, University of P\'ecs,\\ Ifj\'us\'ag \'utja 6, 7624 P\'ecs, Hungary}
\affiliation{ExtreMe Matter Institute EMMI,
   GSI Helmholtzzentrum f\"ur Schwerionenforschung,
   Planckstrasse 1,
   64291 Darmstadt,
   Germany}
\date{\today}

\begin{abstract}
We explore a general framework how to treat coupled-channel systems in the presence of overlapping left and 
right-hand cuts as well as anomalous thresholds. Such systems are studied in terms of a generalized potential, where we exploit the known analytic structure of t- and u-channel forces as the exchange masses get smaller approaching their physical values. Given an approximate generalized potential the coupled-channel reaction amplitudes are defined in terms of non-linear systems of integral equations. For large exchange masses, where there are no anomalous thresholds present, conventional $N/D$ methods are applicable to derive numerical solutions to the latter.  
At a formal level a generalization to the anomalous case is readily formulated by use of suitable contour integrations with amplitudes to be evaluated at complex energies. However,  it is a considerable challenge to  find numerical solutions to anomalous systems set up  on a set of complex contours. By a suitable deformations of left-hand and right-hand cut lines we managed to establish a framework of linear integral equations defined for real energies. Explicit expressions are derived for the driving terms that hold for an arbitrary number of channels. Our approach is illustrated 
in terms of a schematic 3-channel systems. It is demonstrated that despite the presence of anomalous thresholds 
the scattering amplitude can be represented in terms of 3 phase shifts and 3 in-elasticity parameters, as one would expect from the coupled-channel unitarity condition.

\end{abstract}


\maketitle
\tableofcontents

\section{Introduction}

A reliable and systematic treatment of coupled-channel systems subject to strong interactions is still one of the 
remaining fundamental challenges of modern physics. So far effective field-theory approaches with hadronic degrees 
of freedom that reflect QCD properties are established only for particular corners of the strong interaction world. 
At energies where QCD forms bound-states or resonances there is a significant lack of profound theory that connects 
to experimental data directly. Despite the tremendous efforts and successes of experimental accelerator facilities and 
emerging lattice gauge theory simulations there is a significant gap of what theory can do and experimental groups 
would need to be properly guided for new searches of exotic matter \cite{Briceno:2015rlt,Meyer:2015eta,Lutz:2015ejy,Pennington:2016dpj}.  

To unfold the underlying physics of this non-perturbative domain of QCD novel approaches are required that combine the power of 
coupled-channel unitarity together with the micro-causality condition for hadronic degrees of freedom 
\cite{Zachariasen:1962zz,Dalitz:1967fp,Logan:1967zz,Kaiser:1995eg,Oller:1997ng,Oller:1998hw,Nieves:1999bx,GomezNicola:2001as,Lutz:2001yb,Lutz:2001mi,Nieves:2001wt,Kolomeitsev:2003kt,Lutz:2003fm,GarciaRecio:2002td,GarciaRecio:2003ks,Gasparyan:2010xz,Danilkin:2010xd,Danilkin:2011fz,Gasparyan:2012km,Danilkin:2012ua}.
While such frameworks exist for coupled-channel interactions that are dominated by short-range forces matters turn significantly 
more challenging in the presence of t- or u-channel long-range forces \cite{Gasparyan:2010xz,Danilkin:2010xd,Danilkin:2011fz,Gasparyan:2011yw,Gasparyan:2012km,Danilkin:2012ua,Lutz:2011xc,Heo:2014cja,Lutz:2015lca}.  
In particular coupled-channel systems involving the nonet of vector mesons with $J^P=1^-$
or the baryon decuplet states with $J^P = \frac{3}{2}^+$ can only be studied with significant results after such a framework has been developed. 
The latter play a crucial role in the hadrogenesis conjecture that expects the low-lying resonance spectrum of QCD with up, down and strange quarks only, to be generated 
by final state interactions of the lowest SU(3) flavor multiplets with $J^P=0^-, 1^-$ and $J^P = \frac{1}{2}^+,  \frac{3}{2}^+$ \cite{Lutz:2001dr,Lutz:2001yb,Lutz:2001mi,
Kolomeitsev:2003kt,Lutz:2003fm,Lutz:2003jw,Kolomeitsev:2003ac,Lutz:2007sk,Hofmann:2005sw,Hofmann:2006qx,Terschlusen:2012xw}.
The coupled-channel interaction of such degrees of freedom leads to a plethora of subtle effects, like numerous anomalous thresholds  \cite{Mandelstam:1960zz,Ball:Frazer:Nauenberg:1962,Greben:1976wc}. 

The physical relevance of anomalous threshold effects has been discussed recently in \cite{Liu:2015taa,Yang:2016sws,Guo:2016bkl,Liu:2016xly,Xie:2017mbe,Liu:2017vsf}. 
To the best knowledge of the authors there is no established approach available that can treat such phenomena in coupled-channel systems reliably. 
In the previous works which attempted to deal with such systems the strategy was to perform an analytic continuation of an $N/D$ ansatz for the reactions amplitudes in the external mass parameters as 
to smoothly connect a normal system to an anomalous system.  This was studied for two channel systems only \cite{Ball:Frazer:Nauenberg:1962,Greben:1976wc}. Even there 
the first study of Ball, Frazer and Nauenberg \cite{Ball:Frazer:Nauenberg:1962} was rejected by the later work of Greben and Kok as being incorrect \cite{Greben:1976wc}. So far
we have not been able to track any numerical implementation of either of the two  schemes \cite{Ball:Frazer:Nauenberg:1962,Greben:1976wc}. 
Following this strategy an extension to a truly multi-channel systems appears prohibitively cumbersome.

A powerful framework to study coupled-channel systems is based on the concept of
a generalized potential. A partial-wave scattering amplitude  $T_{ab}(s)$ with a channel index $a$ and $b$
for the final and the initial state respectively is decomposed into contributions from left- and right-hand cuts where
all left-hand cut contributions are collected into the generalized potential $U_{ab}(s)$. For a given generalized
potential the right-hand cuts are implied by the non-linear integral equation
\begin{eqnarray}
T_{ab}(s) = U_{ab}(s) +
\sum_{c,d}\int \frac{d \bar s }{\pi} \,\frac{s-\mu^2}{\bar s-\mu^2}\,
\frac{T^\dagger_{ac}(\bar s)\,\hat \rho_{cd}(\bar s)\,T_{db}(\bar s)}{\bar s-s } \,,
\label{def-generalized-potential}
\end{eqnarray}
where $\hat \rho_{cd}(s)$ is a channel dependent phase-space function and all integrals are for $\bar s$ on the real axis.
By construction any solution of (\ref{def-generalized-potential}) does satisfy the coupled-channel
s-channel unitarity condition for normal systems. Typically, the matching scale $\mu^2$ in (\ref{def-generalized-potential}) is to be chosen in a kinematical region where the reaction amplitude can be computed in perturbation theory \cite{Lutz:2001yb,Lutz:2003fm,Gasparyan:2010xz}. 
For normal systems the non-linear and coupled set of equations (\ref{def-generalized-potential}) can be solved by an $N/D$ ansatz
\cite{Chew:1955zz,Mandelstam:1958xc,Chew:1961ev,Frye:1963zz,Ball:1969ri,Chen:1972rq,Eden:1966,Johnson:1979jy,Gasparyan:2010xz,Danilkin:2010xd}. While the general framework is known from the 60's of last
century  only recently
this framework has been successfully integrated into an effective field theory approach based on the chiral
Lagrangian \cite{Gasparyan:2010xz,Danilkin:2010xd,Danilkin:2011fz,Gasparyan:2011yw,Gasparyan:2012km,Danilkin:2012ua,Lutz:2011xc}. 
As it stands (\ref{def-generalized-potential}) this approach breaks down once a coupled-channel system involves unstable particles or 
anomalous thresholds arise. In this work we will construct a suitable adaptation  that overcomes this gap. 

Our formal developments will be illustrated by a schematic three-channel model where explicit numerical results for key quantities will be presented along the way. In Section II and III  we discuss the analytic structure of the coupled-channel reaction amplitudes and propose an efficient deformation of the left- and right-hand cut lines such that anomalous systems can be dealt with. In addition our schematic model is specified. In Section IV an ansatz 
for the solution of the non-linear integral equations on the complex contour lines is derived. In the next step in Section V we consider the limit where the complex contour lines are deformed back towards the real energy line. The main formal result of our work is derived in Section VI, where a set of linear integral equations for anomalous systems that is suitable for a numerical implementations  is established. We close with a summary and outlook in Section VII.

\section{Analytic structure of partial-wave scattering amplitudes }

\begin{figure}[t]
\center{
\includegraphics[keepaspectratio,width=0.8\textwidth]{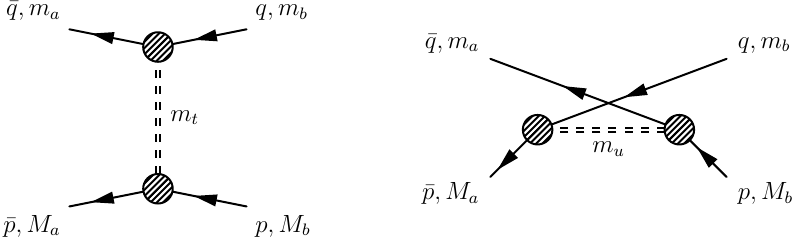} }
\caption{\label{fig:1} Generic t- and u-channel exchange processes.  }
\end{figure}

In a first step we will recall a spectral representation for a generic t-channel and u-channel term as shown in Fig. \ref{fig:1}. 
In our previous work \cite{Lutz:2015lca} we established the following general form 
\begin{eqnarray}
&& U^{(\rm t-ch.)}_{ab} (s)  = \sum_{i=\pm}\,
\int_{-\infty}^{\infty} \frac{d m^2}{\pi}\,\frac{\varrho^{(t)}_{i,ab}(m^2,\,m_t^2)}{s-c^{(t\,)}_{i,ab}(m^2)}\,
\left(\frac{d }{d m^2 }\,c^{(t\,)}_{i,ab}(m^2) \right)\,,
\nonumber\\
&&  U^{(\rm u-ch.)}_{ab} (s)  = \sum_{i=\pm}\,
\int_{-\infty}^{\infty} \frac{d m^2}{\pi}\,\frac{\varrho^{(u)}_{i,ab}(m^2,\,m_u^2)}{s-c^{(u)}_{i,ab }(m^2)}\,
\left(\frac{d }{d m^2 }\,c^{(u)}_{i,ab}(m^2) \right)\,,
\label{disp-general-u-t-channel}
\end{eqnarray}
with properly constructed spectral weights $\varrho^{(t)}_{\pm,ab}(m^2,\,m_t^2)$ and $\varrho^{(u)}_{\pm,ab}(m^2,\,m_u^2)$.
The contour functions $c^{(t)}_{\pm,ab}(m^2)$ and $c^{(u)}_{\pm,ab}(m^2)$ depend on the masses of initial and final particles of the given reaction 
$ab$  for which we use the convenient notation
\begin{eqnarray}
q^2 = m_b^2 \,, \qquad \bar q^2 = m_a^2\,, \qquad p^2= M_b^2 \,,\qquad \bar p^2= M_a^2\,.
\end{eqnarray}
While the derivation of the spectral weights $\varrho^{(t)}_{\pm,ab}(m^2,\,t)$ and $\varrho^{(u)}_{\pm,ab}(m^2,\,u)$
is quite cumbersome the identification of the contour functions $c^{(t)}_\pm(m^2)$ and $c^{(u)}_\pm(m^2)$
is straight forward. Owing to the Landau equations any possible branch point of a partial-wave amplitude
must be associated with an endpoint singularity of the partial-wave projection integral that involves some Legendre polynomials in 
$\cos \theta$ of the scattering angle $\theta$. This leads to the well known result
\begin{eqnarray}
&& c^{(u)}_{\pm,ab}(m^2)  = \frac{1}{2}\,\Big( M_a^2+m_a^2+M_b^2+m_b^2
-m^2\Big)
 +\frac{M_a^2-m_b^2}{\sqrt{2}\,m}\,
\frac{M_b^2-m_a^2}{\sqrt{2}\,m}
\nonumber\\
&&\pm\, \frac{m^2}{2}
\sqrt{ \Bigg(1-2\,\frac{M_a^2+m_b^2}{m^2}
+\frac{(M_a^2-m_b^2)^2}{m^4} \Bigg)\Bigg(
1-2\,\frac{M_b^2+m_a^2}{m^2}
+\frac{(M_b^2-m_a^2)^2}{m^4} \Bigg)}\,,
\nonumber\\ \nonumber\\
&&c^{(t)}_{\pm,ab}(m^2) =  \frac{1}{2}\,\Big(M_a^2+m_a^2+M_b^2+m_b^2-m^2 \Big)
- \frac{M_a^2-M_b^2}{\sqrt{2}\,m}\,
\frac{m_a^2-m_b^2}{\sqrt{2}\,m}
\nonumber\\
&&\pm \,\frac{m^2}{2} \sqrt{\Bigg( 1-2\,\frac{M_a^2+M_b^2}{m^2}
+\frac{(M_a^2-M_b^2)^2}{m^4} \Bigg) \Bigg(
1-2\,\frac{m_a^2+m_b^2}{m^2} +\frac{(m_a^2-m_b^2)^2}{m^4}\Bigg)}\,.
\label{lamtab}
\end{eqnarray}

An anomalous system arises if either of the spectral weights $\varrho^{(t)}_{i}(m^2,\,m_t^2)$ or $\varrho^{(u)}_{i}(m^2,\,m_u^2)$ 
is non-vanishing at an exchange mass $m$ where the associated contour $c^{(t)}_i(m^2)$ or $c^{(u)}_i(m^2)$ approaches any of the thresholds or 
pseudo thresholds of the given reaction $ab$. In this case the decomposition (\ref{def-generalized-potential}) breaks down: an anomalous threshold 
behavior is encountered. The latter is characterized by a particular branch point $ \mu^A_{ab}$ of the partial-wave amplitude 
\begin{eqnarray}
 \mu^A_{ab} < {\rm Min} \{ m_a+M_a , m_b + M_b \}\,,
 \label{def-muA}
\end{eqnarray}
that is associated with the given amplitude $ab$. For simplicity of the presentation we consider in (\ref{def-muA}) an anomalous threshold behavior at a normal threshold point only. A similar phenomenon may occur at a pseudo threshold. 
Also the case $\mu^A_{ab} > {\rm Max} \{ m_a+M_a , m_b + M_b \}$ is not excluded in general.
For both cases, our approach can be adapted in a straight forward manner. 
What is excluded, however, from general considerations, that an anomalous threshold arises for a diagonal reaction with $a=b$.

In the previous works an analytic continuation in the external mass parameters as 
to smoothly connect a normal two-channel system to an anomalous two-channel system was attempted \cite{Ball:Frazer:Nauenberg:1962,Greben:1976wc}. After all using this 
method Mandelstam gave a transparent presentation of the anomalous threshold phenomenon in a typical one-loop diagram \cite{Mandelstam:1960zz}. However, we feel that an application of this method directly to  multi-channel reaction amplitudes, as attempted in \cite{Ball:Frazer:Nauenberg:1962,Greben:1976wc}, appears futile due to the proliferation of branch points and cut structures that need to be properly deformed and followed up in various limits.

\begin{figure}[t]
\vskip-0.1cm
\center{
\includegraphics[keepaspectratio,width=0.95\textwidth]{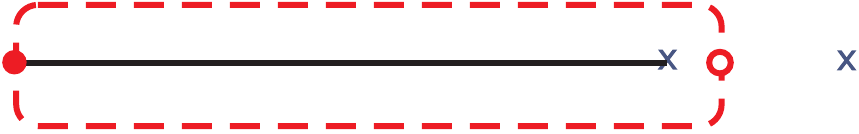} }
\vskip-0.1cm
\caption{\label{fig:2} Left-hand cut line of an anomalous contribution to the generalized potential. The two crosses show the location of the threshold points at $(m_a+M_a)^2$ and 
$(m_b + M_b)^2$ where we assume $a < b$ without loss of generality.
The filled and open red circles indicate the location of the  anomalous threshold point
$(\mu_{ab}^A)^2$ and a return point $(\mu_{ab}^R)^2$ respectively. 
}
\end{figure}

We argue that there is a significantly more transparent path to arrive at a framework that is capable to treat coupled-channel systems in the presence of 
anomalous thresholds. Given a multi-channel system various anomalous thresholds may appear in different reactions. Our starting point, is based directly on the general representation (\ref{disp-general-u-t-channel}), which already established  a suitable analytic continuation of the tree-level potential terms valid for arbitrary exchange masses. 
As was emphasized in our previous work \cite{Lutz:2015lca} the reason why a one-loop diagram develops an anomalous threshold can be read off easily from its driving tree-level 
potential terms. 

We need to be somewhat more specific on the generic form of the generalized potential. The tree-level potential takes the generic form
\begin{eqnarray}
 U^{\rm tree-level}_{ab}(s) = \int_{L_{ab}} \frac{d \bar s}{\pi}\, \frac{\rho^{U}_{ab}(\bar s)}{\bar s- s} \,,
\end{eqnarray}
where the left-hand contour, $L_{ab}$, is a union of contours required for the spectral representation of a general t- or u-channel exchange process as recalled with (\ref{disp-general-u-t-channel}). In order to achieve our goal it is instrumental to separate that left-hand contour into  a normal and an anomalous part
\begin{eqnarray}
 L_{ab} = \Delta L_{ab} \cup A_{ab}  \,,
\end{eqnarray}
where the part $A_{ab}$ starts at the anomalous threshold $ (\mu^A_{ab})^2$ slightly above the real axis, extends to the normal threshold point ${\rm Min } \{(m_a + M_a)^2, (m_b+ M_b)^2 \}$ and returns at 
$(\mu_{ab}^R)^2$ to the anomalous one slightly below the real axis. This is illustrated in Fig. \ref{fig:2}. Note that according to the general representation (\ref{disp-general-u-t-channel}) 
the anomalous contour line would be on real axis only. With Fig. \ref{fig:2} such a contour is  deformed into the complex plane in a manner that leaves its contribution to the generalized  potential unchanged for $s$ outside the area encircled by the dashed line. 
While the values for $\mu_{ab}^A$ are determined by the specifics of the coupeld-channels interactions, there is some freedom how to choose the location of the return points $\mu_{ab}^R$. Given the analytic structure of the spectral weight $\rho^{U}_{ab}(\bar s)$ the generalized potential $U^{\rm tree-level}_{ab}(s)$ does not depend on the choice of the return point $\mu^R_{ab}$ at energies $s$ outside the dashed line of Fig. \ref{fig:2}.

We point at a subtle issue such a contour separation is based on. Here we tacitly assumed that the anomalous threshold $\mu^A_{ab}$ is real. However, from the general representation (\ref{disp-general-u-t-channel}) there is the possibility of a pair of complex conjugate anomalous points $\mu^{A\pm}_{ab}$. In this case the associated  contour lines have to be followed till both reach the common pseudo-threshold point at $s = (\mu^A_{ab})^2={\rm Max }\{(m_a - M_a)^2, (m_b- M_b)^2\}$. The latter we take as the anomalous threshold value in our work. This can be justified since in this case  the spectral weight $\rho^{ U}_{ab}(\bar s)$ can be shown to be analytic an $\epsilon$ left to that pseudo threshold point at $\bar s = (\mu^A_{ab})^2- \epsilon$.

It is useful to identify an 
anomalous threshold value  $\mu_a$ associated with a given channel $a$. For the clarity of the development we assume in the following a strict channel ordering 
according to the nominal threshold value, i.e. we insist on
\begin{eqnarray}
m_{a} + M_{a} \leq m_{a+1} + M_{a+1} \qquad {\rm for\; all} \qquad a\,,
\label{threshold-ordering}
\end{eqnarray}
where for later convenience we permit the equal sign for channels with the same two-particle states in different spin configurations only. 
We can now identify the desired anomalous threshold value.
It is the minimum of all accessible anomalous branch points
\begin{eqnarray}
\mu_a \equiv \underbrace{{\rm Min}}_{b> a}\,\Big\{\mu^A_{ab},m_a + M_a\Big\}\,, \qquad \qquad \qquad 
\label{def-anomalous-threshold}
\end{eqnarray}
where  the value $\mu^A_{ab}$ gives the anomalous threshold value of the partial-wave amplitude $ab$. Note that it does not necessarily follow that 
the channel ordering (\ref{threshold-ordering}) implies $\mu_{a+1} \geq \mu_a$. If such a channel crossing with $\mu_{a} > \mu_b$ for any pair ab with $a <b$ occurs, we will further 
move the point $\mu_a$ with $\mu_a \to \mu_b -\epsilon $ such that we ultimately arrive at a  strict channel ordering
\begin{eqnarray}
 \mu_{a }\leq   \mu_{a+1} \qquad {\rm for\; all} \qquad a\,,
\end{eqnarray}
where again the equal sign is permitted for channels with the same two-particle states in different spin configurations only. It is useful to introduce a similar streamline 
of the plethora of return points. We choose universal return points $\hat \mu_a$ of the particular form 
\begin{eqnarray}
 \mu^R_{ab} = \hat \mu_a < m_b + M_b \,\qquad {\rm for\; all} \qquad b> a \,,
\end{eqnarray}
where we consider $a$ and $b$ with $\mu^A_{ab} < (m_a + M_a)^2 $ only.

We can now introduce our anomalous contour $A_a$ that starts at $\mu_a^2$ slightly above the real axis, passes $(m_a+ M_a)^2$ and returns at $\hat \mu_a^2$ to $\mu_a^2$ below the real axis. In turn we can now write
\begin{eqnarray}
 U_{ab}(s) = \int_{\Delta L_{ab}} \frac{d \bar s}{\pi}\, \frac{\rho^{\rm normal}_{ab}(\bar s)}{\bar s- s} + \int_{A_{{\rm Min} (a,b)}} \frac{d \bar s}{\pi}\, \frac{\rho^{\rm anomalous}_{ab}(\bar s)}{\bar s- s} + \cdots\,,
 \label{res-Urho}
\end{eqnarray}
where we will exploit that the first term in (\ref{res-Urho}) is uncritical to the extent that it is analytic in the vicinity of any normal or anomalous threshold point. The challenging term is the second one, which is associated
with a contour that entangles the normal threshold point ${\rm Min} \{(m_a+ M_b)^2, (m_b + M_b)^2 \}$. The dots in (\ref{res-Urho}) remind us that we did not yet discuss the generic form of contributions to the generalized potential from loop effects.

\begin{figure}[t]
\vskip-0.1cm
\center{
\includegraphics[keepaspectratio,width=0.95\textwidth]{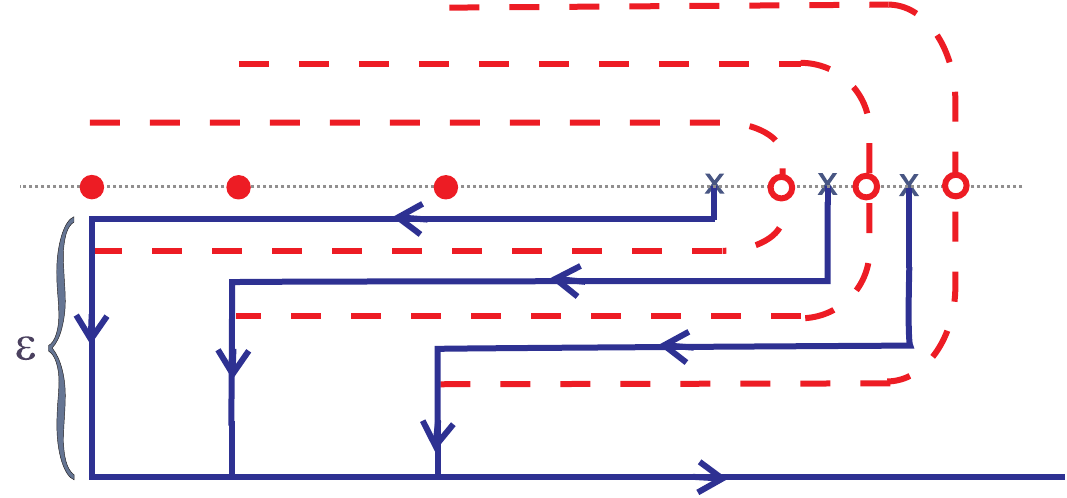} }
\vskip-0.1cm
\caption{\label{fig:3} Deformed  left- and right hand cut lines for the reaction amplitudes. 
The crosses show the location of the threshold points at $(m_a+M_a)^2$. While the anomalous left-hand cut lines are shown with  dashed red lines, the deformed right-hand cut lines are represented by blue solid lines. The filled and open red circles indicate the location of the  anomalous threshold points $\mu_a^2$ and the return points $\hat \mu_a^2$  respectively. }
\end{figure}

The key starting point is an adaptation of the non-linear integral equation (\ref{def-generalized-potential}). The integral in (\ref{def-generalized-potential})
starts at the point where the phase-space matrix $\rho_{cd}(\bar s)$ vanishes at $\bar s= (m_c + M_c)^2= (m_d + M_d)^2$ and remains on the real axis going to infinity. 
In the anomalous case the integral must also start at $\bar s= (m_c + M_c)^2= (m_d + M_d)^2$ but will leave the real axis following suitable paths 
on higher Riemann sheets. The integral on the real axis has to be replaced by a contour integral. 
Eventually, somewhere on its way the contour path will touch the anomalous threshold at $\bar s = \mu^2_c$. 
Note that the case $c\neq d$ appears only for systems with non-vanishing spin. 

The reason for this complication is a consequence of the anomalous threshold behavior of the generalized potential. If we wish to separate the left- from the right-hand cuts 
in the reaction amplitudes, it is necessary to apply suitable deformations of the cut-lines. In the normal case the separation of left- from right-hand cuts is 
trivially implied by (\ref{def-generalized-potential}). The second term on the right-hand-side has a cut on the real axis extending from $\bar s= (m_1+M_1)^2$ to infinity. 
This is the right-hand cut. All left-hand cuts sit in $U(s)$. As long as the cut lines of the generalized potential do not cross any of the right-hand cut lines 
on the real axis the non-linear integral equation (\ref{def-generalized-potential}) is well defined and can be solved numerically in application of conventional methods. This is not the case if the generalized 
potential develops  an anomalous threshold. Without an appropriate adaptation the associated left-hand cut lines would cross the right-hand cut-lines. An avoidance is possible only 
if the integral on the real lines in (\ref{def-generalized-potential}) is replaced by contour integrals in the complex plane.

In Fig. \ref{fig:3} we illustrate the analytic structure of the reaction amplitudes defined with respect to suitably deformed left- and right-hand cut lines. The dashed red lines show the location of the anomalous left-hand cuts. We suggest a deformation of the right-hand cut lines as illustrated by the solid lines in the figure. 
Now, none of the anomalous left-hand cut lines $A_a$ crosses any of the right-hand contour lines $C_a$.

Before writing down the generalization of (\ref{def-generalized-potential})
we need some more notation. The reaction amplitudes $T_{ab}(s)$ need to be evaluated slightly below and above any of the many different contour lines. With
\begin{eqnarray}
 T^{c\pm}_{ab}(\bar s) = T_{ab}(s_\pm  ) \qquad {\rm with} \qquad \bar s \in C_c \,,
 \label{def-Apm}
\end{eqnarray}
where the points $s_+ $ and $s_-$ are an $\epsilon'$ distance above and below the contour $C_c $ at $\bar s$ respectively. 
The value of  $\epsilon'$ is chosen smaller than the minimal distance of any of the horizontal lines in Fig. \ref{fig:3}.

With this notation we can write down the desired generalization
\begin{eqnarray}
T_{ab}(s) = U_{ab}(s) +
\sum_{c,d}\int_{C_c } \frac{d \bar s }{\pi} \,\frac{s-\mu^2}{\bar s-\mu^2}\,
\frac{T^{c-}_{ac}(\bar s)\,\rho_{cd}(\bar s)\,T^{c+}_{db}(\bar s)}{\bar s-s} \,,
\label{def-anamalous-potential}
\end{eqnarray}
where $C_c=C_d$ holds by construction.
We consider the phase-space matrix $\rho_{cd}(s)$ to be analytic on the real axis 
at $s> (m_c+M_c)^2$. This implies that at $s< (m_c+M_c)^2$ the function $\rho_{cd}(s)$ has a cut line.  Note that none of the contour paths $C_c$ cross any of the 
cut-lines of the phase-space functions. It is emphasized that in the absence of an anomalous threshold in the generalized potential our generalization (\ref{def-anamalous-potential}) 
reproduces the conventional expression (\ref{def-generalized-potential}).

\begin{figure}[t]
\center{
\includegraphics[keepaspectratio,width=0.95\textwidth]{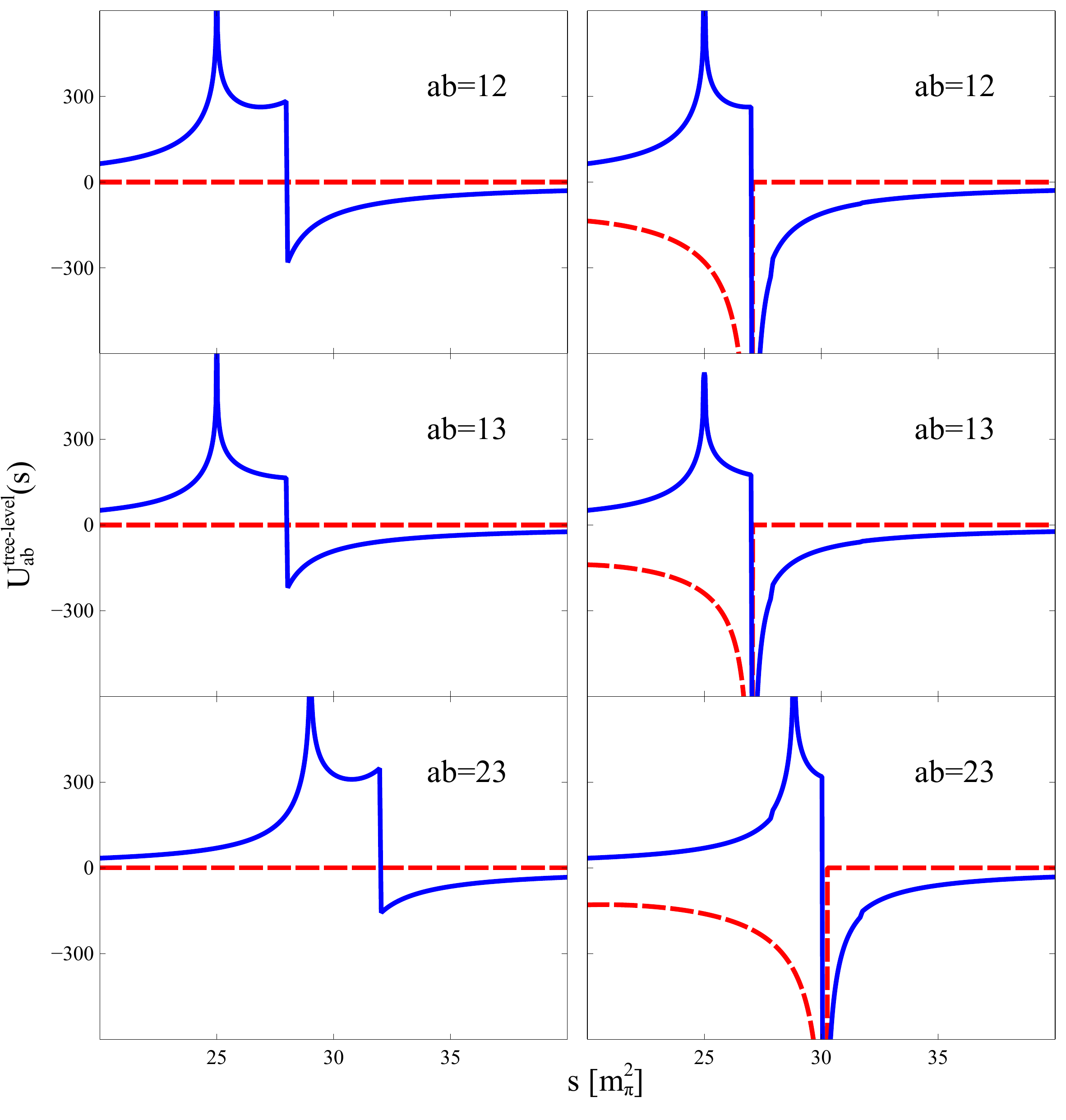} }
\caption{\label{fig:4} The generalized potential $U^{\rm tree-level}_{ab}(s)$ (l.h.p.) and $U^{\rm tree-level}_{ab}(s-i\,\epsilon)$ (r.h.p.) in the schematic model as defined in (\ref{def-model}). Only non-vanishing elements are considered. Real and imaginary parts are shown with solid blue and dashed red lines respectively. }
\end{figure}

\begin{figure}[t]
\center{
\includegraphics[keepaspectratio,width=0.95\textwidth]{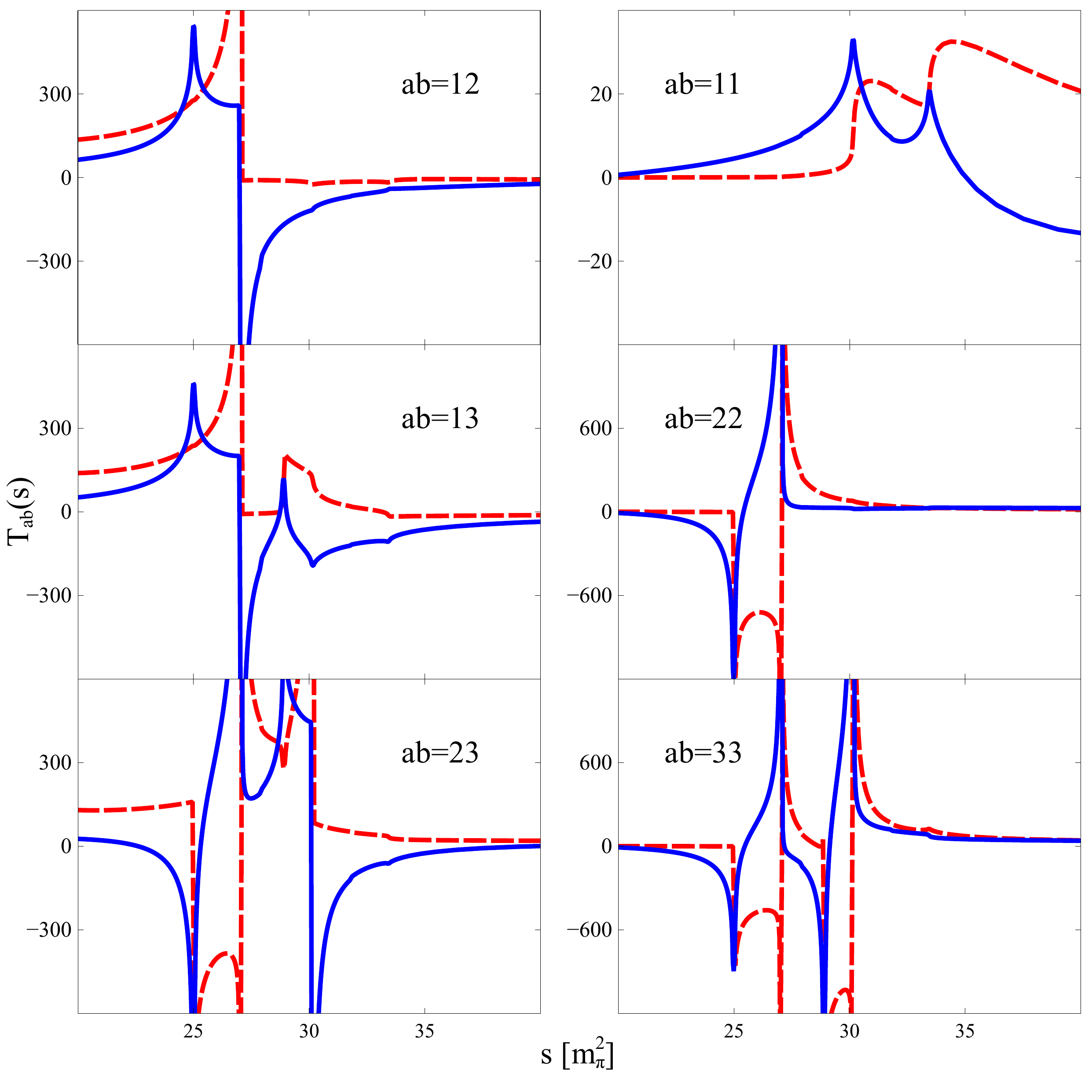} }
\caption{\label{fig:5} Reaction amplitudes $T_{ab}(s + i\,\epsilon)=T_{ba}(s + i\,\epsilon)$ in the schematic model of (\ref{def-model}). Real and imaginary parts are shown with solid blue and dashed red lines respectively.}
\end{figure}

The representation (\ref{def-anamalous-potential}) appears deceivingly simple and ad-hoc, however, we derived it  by an analytic continuations in the t- and u-channel 
exchange mass parameters. The starting point for this is provided by the general representation (\ref{disp-general-u-t-channel}). First we assume all exchange particles to have a very large mass, so that all left-hand branch cuts are well separated from the right-hand branch cuts. Already at this stage we can deform the right-hand cuts from their conventional choice to take the form as indicated in Fig. \ref{fig:3} by the solid lines. This contour deformation cannot change the value of any of the coupled-channel reaction amplitudes $T_{ab}(s)$ at $ \Im s > 0$, simply because along the cut deformations the reaction amplitudes are analytic by assumption. Here we exploit the fact that in the limit of very large exchange masses all left-hand branch cuts are moved outside the figure. In a second step we follow the left-hand cuts as they move right in Fig. \ref{fig:3} with the 
exchange masses getting smaller approaching their physical values \cite{Mandelstam:1960zz,Ball:Frazer:Nauenberg:1962,Frye:1963zz,Greben:1976wc,Johnson:1979jy}. At the end we claim that they can be presented by the  dashed lines in the figure. In turn we take (\ref{def-anamalous-potential}) as a transparent starting point of our presentation.

Our numerical examples will 
rest on a minimal model as it is implied by schematic tree-level interactions.
Along the formal derivations in the next few sections we will illustrate the important auxiliary  quantities in terms of a schematic  three-channel model specified 
with
\allowdisplaybreaks[1]
\begin{eqnarray}
&& \rho_{ab}(s)= \frac{p^+_{a}(s)}{8\,\pi\sqrt{s}}\,\delta_{ab}\,,\qquad \qquad \qquad \qquad \quad 
p^\pm_a =\sqrt{\pm \Big(s-(m_a+M_a)^2\Big)\,\frac{s-(m_a-M_a)^2}{4\,s} }\,,
\nonumber\\
&& \rho_{ab}^{A} (s) = \frac{20\,m_\pi^2}{p^{\,+}_a(s)\,i\,p^-_b(s)}\,\Theta \Big[s-\mu_a^2 \Big]\,\Theta\Big[ \hat \mu_a ^2- s \Big]= \rho_{ba}^{A} (s) \,\qquad {\rm for }\qquad a < b\,,
\nonumber\\
&&  U^{\rm model}_{ab}(s) =  \int_{A_{{\rm Min} (a,b)}} \frac{d \bar s}{\pi}\, \frac{\rho^{A}_{ab}(\bar s)}{\bar s- s} \,,
\nonumber\\ \label{def-model}\\
 && m_1 =  m_2 = m_3 = m_\pi\,,\qquad \qquad \qquad \qquad \quad \mu^2 = 20\,m_\pi^2\,,
 \nonumber\\
 && M_1 = 4.2\,m_\pi \,, \qquad \; M_2 = 4.5\,m_\pi \,,\qquad \qquad  M_3 = 4.8\,m_\pi\,,
 \nonumber\\
 && \mu^2_1 = 25\,m_\pi^2\,, \qquad \quad \; \hat \mu_1 ^2 = 28\,m_\pi^2\,, 
 \qquad \qquad \;\; \mu_2^2 = 29\,m_\pi^2\,,\qquad \qquad   \;\, \hat \mu_2 ^2 = 32 \,m_\pi^2\,. \nonumber
 \end{eqnarray}
In Fig. \ref{fig:4} we plot the tree-level generalized potential for two kinematical cases. The left-hand panel shows the non-vanishing elements with $s$ on the real axis strictly. Here the potentials are characterized by significant variations close to the anomalous threshold points with $s =\mu_a^2$ and at the return points with $s = \hat \mu_a^2$ and 
$\mu_a^2 < (m_a+M_a)^2 < \hat \mu_a ^2$. The potentials are strictly real always and smooth at the threshold points $s = (m_a+M_a)^2$. 
The right-hand panel shows the corresponding potentials evaluated at $ s -i\,\epsilon$, just below the double cut structures as illustrated in Fig. \ref{fig:2}. In this  case the potentials have an imaginary part and in addition are singular at the threshold points $s = (m_a+M_a)^2$. 

In Fig. \ref{fig:5} we anticipate the usefulness of our formal developments and present the solution to the non-linear integral equation (\ref{def-anamalous-potential}) as implied by  (\ref{def-model}).  It is noted that our results do not depend on the particular choices for the return points $\hat \mu_a$ in (\ref{def-model}).
The reaction amplitudes show significant and non-trivial structures that are implied by the presence of the anomalous threshold effects in the generalized potential. While the amplitudes at subthreshold energies show various singular structures, they are smooth and well behaved in the physical region. We emphasize that the subthreshold structures are not driven exclusively by the contribution from the generalized potential as shown in Fig. \ref{fig:4}. There are in addition significant structures generated by the right-hand cut contributions. 
As to the best knowledge of the authors with Fig. \ref{fig:5} we encounter the 
first numerical solution of such an anomalous 3-channel system in the published literature. 

\section{Anomalous thresholds  and coupled-channel unitarity}

Before we explain how to find numerical solutions to the non-linear integral equations (\ref{def-anamalous-potential}) in the presence of anomalous threshold effects it is useful to pause and discuss a critical  issue 
that we will be confronted with. Given a specific approximation for the generalized potential, a solution of (\ref{def-anamalous-potential}) does not necessarily imply that 
the coupled-channel unitarity condition is fulfilled. From the latter we expect a representation of the reaction amplitudes $T_{ab}(s) $ in terms of a set of real quantities, channel dependent phase shift and 
inelasticity parameters $\phi_a(s)$ and $\eta_a(s)$. It should hold
\begin{eqnarray}
&& \Im T_{ab}(s +i\,\epsilon) = \sum_{c,d}\,T^*_{ca}(s +i\,\epsilon)\,\rho_{cd}(s)\,T_{db}(s+i\,\epsilon)\,\Theta \big[ s- (m_c+ M_c)^2\big]\,\qquad\! 
\nonumber\\
&& \qquad \qquad \quad    {\rm for}\qquad \!s > {\rm Max } \{ (m_a + M_a)^2, (m_b + M_b)^2 \} \,,
\nonumber\\ \nonumber\\
&& \sum_c T_{ac}(s) \, \rho_{ca}(s)= \frac{1}{2\,i}\,\Big[ \eta_a(s) \,e^{2\,i\, \phi_a(s)} - 1 \Big] \,, \qquad 
 \label{def-unitarity}
\end{eqnarray}
where on general grounds one expects $0 \leq \eta_a \leq 1$. 
While any solution to (\ref{def-anamalous-potential}) satisfies the time-reversal invariance condition $T_{ab}(s) = T_{ba}(s)$, the Schwarz reflection principle, 
\begin{eqnarray}
T^*_{ab}(s) = T_{ab}(s^*) \,,
\label{def-Schwarz}
\end{eqnarray}
cannot be derived in general for right-hand cut lines off the real axis. It is not surprising that then the coupled-channel unitarity condition is not necessarily obtained.

Let us be specific and identify the generalized potential by tree-level t- and u-channel exchange processes, 
a typical strategy in hadron physics. Though we may solve the non-linear system 
(\ref{def-anamalous-potential}) in this case, the unitarity condition (\ref{def-anamalous-potential}) will not be 
fulfilled once an anomalous threshold  effect is encountered. A supposedly related problem was noted in the previous studies \cite{Ball:Frazer:Nauenberg:1962,Greben:1976wc}.  It was argued that the problem is caused by the neglect of second order contributions to the generalized potential. A minimal ansatz for a physical approximation to the generalized potential requires  some additional terms 
 \begin{eqnarray}
  U_{ab}(s) = U^{\rm tree-level}_{ab}(s) +  U^{\rm box}_{ab}(s )\,,
  \label{res-anomalous}
 \end{eqnarray}
to be constructed properly. However, no conclusive 
form of the latter was presented and illustrated in the literature so far. Conflicting suggestions were put forward in  \cite{Ball:Frazer:Nauenberg:1962,Greben:1976wc} for two-channel systems. Whether any of such forms lead to physical results remains an open challenge. So far there is no numerical implementation for a specific example worked out, at hand of which one may judge the significance of any of the two approaches. Unfortunately, for us both works  \cite{Ball:Frazer:Nauenberg:1962,Greben:1976wc} are rather difficult to follow. 
Nevertheless, we tend to agree with the conclusions of \cite{Greben:1976wc} that the approach advocated in 
\cite{Ball:Frazer:Nauenberg:1962} is incorrect.

\begin{figure}[t]
\center{
\includegraphics[keepaspectratio,width=0.95\textwidth]{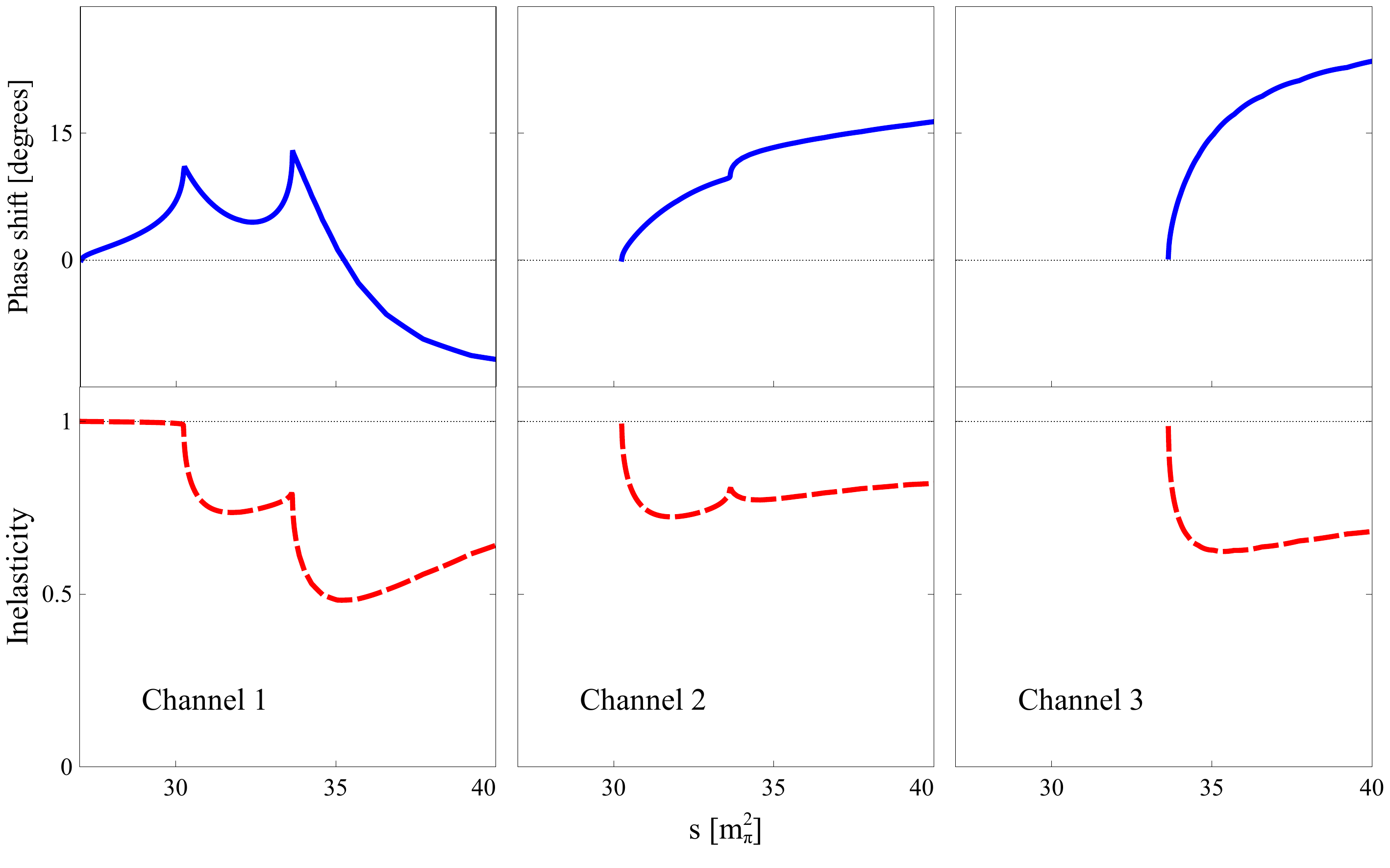} }
\caption{\label{fig:6} Phase shifts and in-elasticity parameters of (\ref{def-unitarity}) in our schematic model (\ref{def-model}).  }
\end{figure}

An important achievement of the current work is the construction of a minimal second order term applicable for systems of arbitrary high dimensions. Our approach is based  on the request that we arrive at the coupled-channel unitarity condition (\ref{def-unitarity}) and recover the Schwarz reflection principle (\ref{def-Schwarz}). The key observation is that in the absence of the anomalous box term $ U^{\rm box}_{ab}(s )$ the second order reaction amplitude $T_{ab}(s)$, as it would follow from (\ref{def-anamalous-potential}), is at odds with  (\ref{def-Schwarz}). The extra term is unambiguously determined by the condition that its inclusion 
restores the Schwarz reflection principle (\ref{def-Schwarz}). This leads to the following form
 \begin{eqnarray}
&& U^{\rm tree-level}_{ab}(s) = \int_{\Delta L_{ab}} \frac{d \bar s}{\pi}\, \frac{\rho^{\rm tree-level}_{ab}(\bar s)}{\bar s- s} + \int_{A_{{\rm Min} (a,b)}} \frac{d \bar s}{\pi}\, \frac{\rho^{A}_{ab}(\bar s)}{\bar s- s} \,,
\nonumber\\
&& U^{\rm box}_{ab}(s ) = -2\,\sum_{c,d < {\rm Min}(a,b)}\int_{A_c}\,\frac{d \bar s}{\pi}\,\frac{ s - \mu_M^2}{\bar s- \mu_M^2}\,\frac{\rho^{A}_{ac}(\bar s)\,\rho_{cd}(\bar s)\,\rho^{A}_{cb}(\bar s)}{\bar s- s}\,,
\label{def-Ubox}
 \end{eqnarray}
 where the spectral weight $\rho^{A}_{ab}(\bar s)$ characterizes the anomalous behavior of the 
 tree-level potential.  It is identified in analogy to the general representation (\ref{res-Urho}).
 Unfortunately, a direct comparison of our result with the ansatz in \cite{Ball:Frazer:Nauenberg:1962,Greben:1976wc} is not so easily possible. Nevertheless, we note that for the two-channel case our result (\ref{def-Ubox}) appears quite compatible with the ansatz discussed in \cite{Greben:1976wc}.  
 
 We anticipate the phase shifts  and in-elasticities as implied by our schematic model (\ref{def-model}) as properly supplemented by the anomalous box term (\ref{res-anomalous}, \ref{def-Ubox}). We affirm that the reaction amplitudes as already shown in Fig. \ref{fig:5} are compatible with the coupled-channel unitarity condition (\ref{def-unitarity}) and   
 the phase shifts and in-elasticity parameters of Fig. \ref{fig:6}. Without an explicit computation the authors would not have been in a position to even roughly guess the non-trivial behavior seen in that figure. 
 
 \begin{figure}[t]
\center{
\includegraphics[keepaspectratio,width=0.95\textwidth]{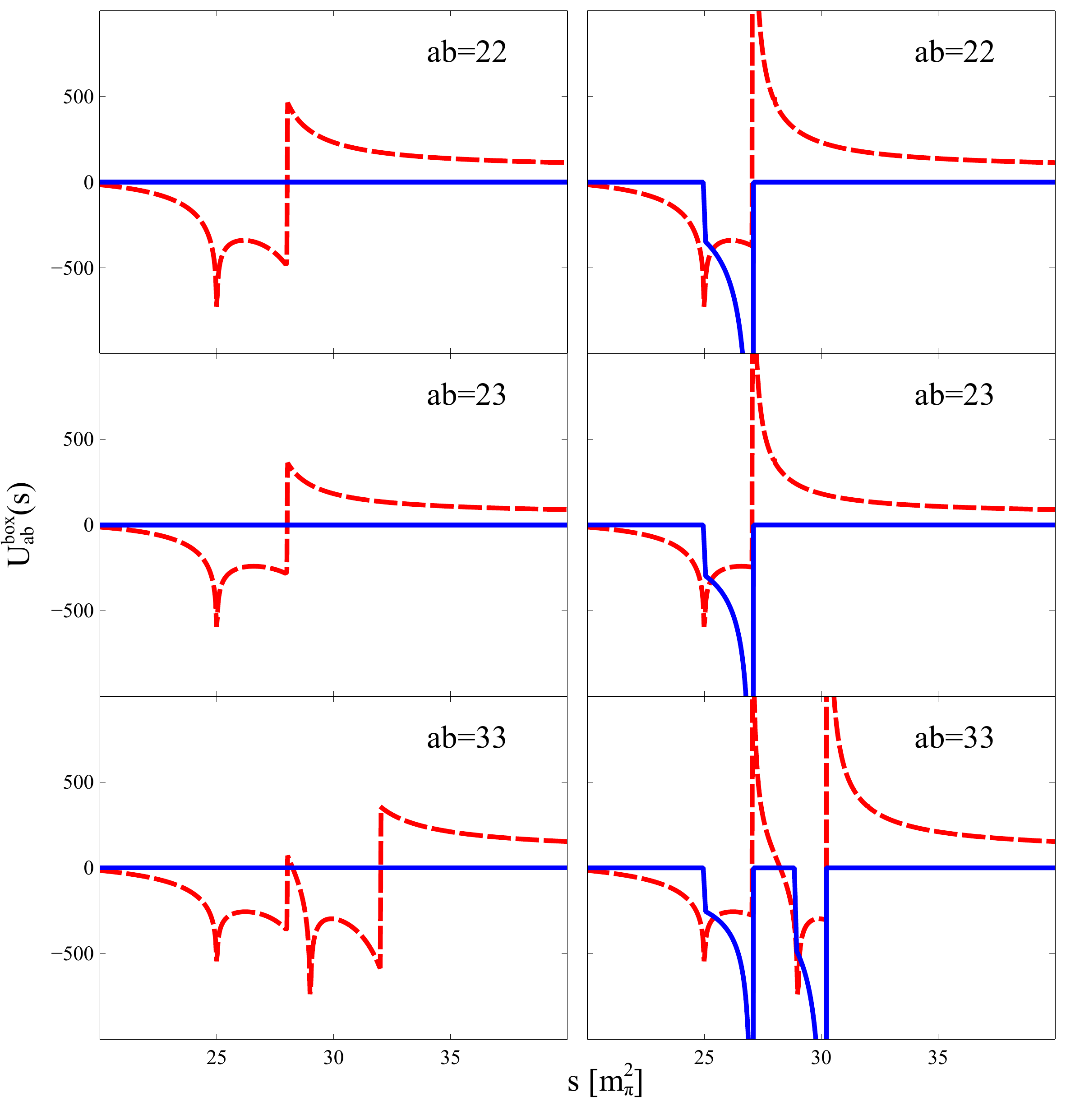} }
\caption{\label{fig:7} The box contributions $U^{\rm box}_{ab}(s)$ (l.h.p.) and 
$U^{\rm box}_{ab}(s-i\,\epsilon)$ (r.h.p.) as defined in (\ref{def-Ubox}) with our model input (\ref{def-model}). Only non-vanishing elements are considered. 
Real and imaginary parts are shown with solid blue and dashed red lines respectively.}
\end{figure}
 
 It is instrumental to realize the quite different nature of the two contributions in (\ref{res-anomalous}). Consider first the tree-level term  in (\ref{res-anomalous}, \ref{def-Ubox}). The spectral weight $\rho_{ab}^{A}( s)$ is real as $s$ approaches the real line below $ s < {\rm Min} \{ (m_a+ M_a)^2,\, (m_b + M_b )^2\}$. 
 In contrast it is purely imaginary for $s > {\rm Min} \{( m_a+ M_a)^2,\, (m_b + M_b )^2\}$. This follows 
 from the general results presented in \cite{Lutz:2015lca}. 
 The corresponding generalized potential satisfies the Schwarz reflection principle with 
  \begin{eqnarray}
  \big[ U^{\rm  tree-level}_{ab}(s) \big]^* = U^{ \rm  tree-level}_{ab}(s^*) \,.
  \label{res-Utree}
 \end{eqnarray}
 We turn to the second order term in (\ref{res-anomalous}, \ref{def-Ubox}).  
 While  for $\Re s <(m_c + M_c)^2 =(m_d + M_d)^2 $ the two factors, $\rho^{A}_{ac}( s)$ and $\rho^{A}_{cb}(s)$, are real quantities as $ s$ approaches the real line, the phase-space factor $\rho_{cd}(s )$ turns purely imaginary in this case. This leads to the property
 \begin{eqnarray}
   \big[ U^{\rm box}_{ab}(s) \big]^* = - U^{ \rm box}_{ab}(s^*)\,, \qquad \qquad 
  \label{res-Ubox}
 \end{eqnarray}
and illustrates the particular feature of the anomalous box term our construction is based on.  Given our schematic model (\ref{def-model}) the anomalous box term $U_{ab}^{\rm box}(s)$ is illustrated with Fig.\ref{fig:7}. Like in the previous Fig. \ref{fig:6} the left-hand panel shows the non-vanishing potentials on the real axis, the right-hand panel the corresponding potentials slight below the real axis at $s -i\,\epsilon$.

It should be emphasized that  there are further second order contributions to the generalized potential, however, 
they add to the 'normal' spectral weight $\rho_{ab}^{ \rm normal}(\bar s)$ in (\ref{res-Urho}) only. Since the latter terms do not jeopardize the coupled-channel unitarity condition, there is no stringent reason to consider such effects in an initial computation. Typically one may hope that the effect of the latter is suppressed in some suitable power-counting scheme. This should be so since higher loop effects are characterized by left-hand branch cuts that are further separated from the right-hand cuts. In turn such contributions to the generalized potential 
cannot show any significant variations at energies where the generalized potential is needed in (\ref{def-anamalous-potential}).

We would speculate, that the ansatz (\ref{res-anomalous}) is quite generic, i.e. it should hold also for contributions including higher loop effects 
\begin{eqnarray}
 && U_{ab}(s) = \int_{\Delta L_{ab}} \frac{d \bar s}{\pi}\, \frac{\rho^{\rm normal}_{ab}(\bar s)}{\bar s- s} + \int_{A_{{\rm Min} (a,b)}} \frac{d \bar s}{\pi}\, \frac{\rho^{\rm anomalous}_{ab}(\bar s)}{\bar s- s} 
 \nonumber\\
&& \qquad \quad  + \sum_{c < {\rm Min} (a,b) } \int_{A_c} \frac{d \bar s}{\pi}\, \frac{\rho^{\rm anomalous}_{ab,c}(\bar s)}{\bar s- s}\,,
 \label{res-Urho-conjecture}
\end{eqnarray}
where we expect $\rho^{\rm anomalous}_{ab,c}(\bar s)$ to be determined by  $\rho^{\rm anomalous}_{ac}(\bar s)$ and $\rho^{\rm anomalous}_{cb}(\bar s)$ in analogy to (\ref{def-Ubox}).

\section{Non-linear integral equation on complex contours}

The key issue is how to numerically solve that non-linear set of equation (\ref{def-anamalous-potential}) and cross check its physical correctness. After all one may consider 
it merely as a definition of the generalized potential in the presence of anomalous thresholds. For a given approximated generalized potential $U_{ab}(s)$ 
we will device an appropriate $N/D$ like ansatz that will eventually lead to a framework which is amenable to numerical simulations of (\ref{def-anamalous-potential}). 

\begin{figure}[t]
\center{
\includegraphics[keepaspectratio,width=0.95\textwidth]{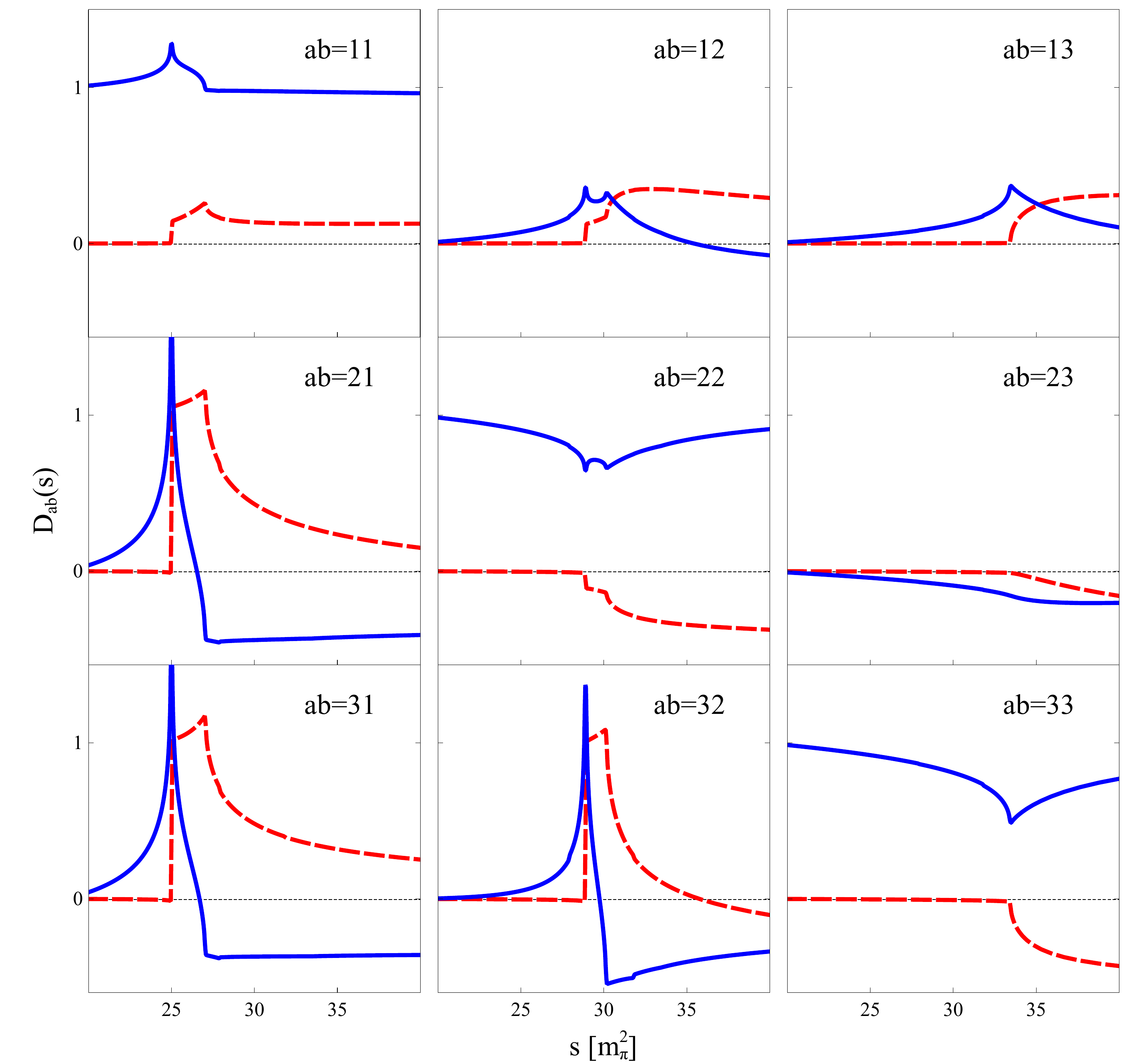} }
\caption{\label{fig:8} The functions $D_{ab}(s+ i\,\epsilon)$ of (\ref{def-varsigma}) in our schematic model (\ref{def-model}). Real and imaginary parts are shown with solid blue and dashed red lines respectively.   }
\end{figure}

We introduce a set of contour functions $\varsigma_{ab}(\bar s)$ defined initially on distinct contours $C_b$, the choice of which depends on the 
channel index $b$ 
\begin{eqnarray}
&& \varsigma_{ab}(\bar s) 
= - \sum_{c,d} D^{b+}_{ac}(\bar s)\, T^{b+}_{cd}(\bar s)\,\rho_{db}(\bar s)  
= - \sum_{c,d} D^{b-}_{ac}(\bar s)\, T^{b-}_{cd}(\bar s)\,\rho_{db}(\bar s) \,
\qquad {\rm for} \qquad 
\bar s \in C_b \,,
\nonumber\\
&& D_{ab}(s) =\delta_{ab} + \int_{C_b} \frac{d \bar s }{\pi} \,\frac{s-\mu^2}{\bar s-\mu^2}\,
\frac{\varsigma_{ab}(\bar s)}{\bar s-s } \,,
\label{def-varsigma}
\end{eqnarray}
where we apply the convenient $\pm$ notation introduced already in (\ref{def-Apm}). 
Assuming the existence of such a set of functions 
$\varsigma_{ab}(\bar s)$ we seek to express the reaction amplitude $T_{ab}( s)$ in terms of them.
This requires a few steps. Like in the previous sections we anticipate with Fig. \ref{fig:8} the form of the $D$ functions as they are implied in our schematic model.
This may help the reader to fight through the various abstract arguments presented in the following.  It is emphasized that none of the function $D_{ab}(s + i\,\epsilon)$ depend on the particular choice of the return points $\hat \mu_{ab}$ in (\ref{def-model}). The functions have a significant imaginary part starting at the anomalous threshold $s \geq \mu^2_b$.

\begin{figure}[t]
\center{
\includegraphics[keepaspectratio,width=0.95\textwidth]{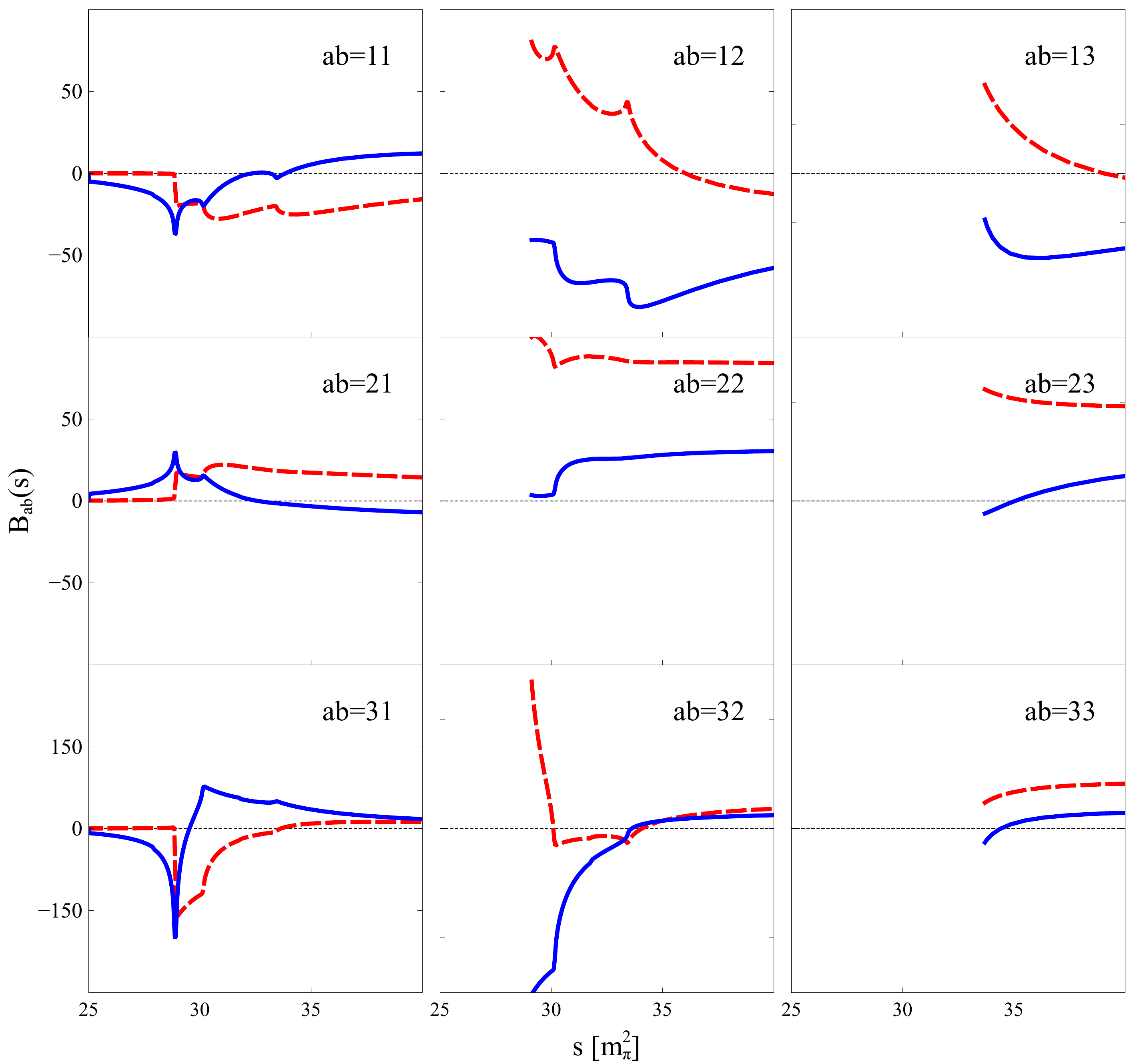} }
\caption{\label{fig:9} The functions $B_{ab}(s+ i\,\epsilon)$ of (\ref{def-B-function}) in our schematic model (\ref{def-model})  for $s> \mu_b^2$ or $s> (m_b + M_b)^2$. Real and imaginary parts are shown with solid blue and dashed red lines respectively.   }
\end{figure}

In the first step we study the analytic properties of the function 
\begin{eqnarray}
&& B_{ab}(s) \equiv \sum_{d} D_{ad}(s)\,\Big[  U_{db}(s) - T_{db}(s)\Big] 
\nonumber\\
&& \qquad \quad =-\sum_{c,d,e}  D_{ad}(s)\, 
\int_{C_c} \frac{d \bar s }{\pi} \,\frac{s-\mu^2}{\bar s-\mu^2}\,
\frac{T^{c-}_{dc}(\bar s)\,\rho_{ce}(\bar s)\,T^{c+}_{eb}(\bar s) }{\bar s-s } \,,
\label{def-B-function}
\end{eqnarray}
where we applied the master equation (\ref{def-anamalous-potential}). 
In Fig. \ref{fig:9} we provide such functions $B_{ab}(s + i\,\epsilon)$ as implied by our model (\ref{def-model}) for $s > (m_b+ M_b)^2$, a region which encompasses the physical domain probed by the phase shifts and in-elasticity parameters in (\ref{def-unitarity}). Here we again do not encounter any dependence on the return points $\hat \mu_a$. However, the functions $B_{a1}(s)$ show a strong variation close to the anomalous threshold point at $s = \mu_2^2 = 29 \,m_\pi^2$. It is important to note that this is not propagated into the amplitude $T_{11}(s+ i\,\epsilon)$ as is evident with Fig. \ref{fig:5} and Fig. \ref{fig:6}.

From (\ref{def-varsigma}) it follows that the function $D_{ab}(s)$ is analytic in the complex plane 
with the exception of a cut along the contour $C_b$. A similar conclusion can be drawn for the 
function $B_{ab}(s)$ only that in this case the cut line is a superposition of all right-hand contours $C_{c}$. 
Since the contours $C_c$ partially overlap it is useful to decompose the contours with 
\begin{eqnarray}
 C_c = C^+_c + C_c^- \qquad {\rm and } \qquad C_A = C^-_1 
\qquad \to    \qquad \sum_c C_c = C_A + \sum_c  C^+_c\,,
\end{eqnarray}
where the lower $C^-_c$ contours are all on the straight line crossing the value  $s =-i\,\epsilon$. 
By construction the upper contours $C^+_c$ do not overlap. 

From this we conclude that a dispersion 
integral representation of the form 
\begin{eqnarray}
B_{ab}(s) &=&\frac{1}{2\,i} \sum_c\int_{ C^+_c} \frac{d \bar s }{\pi} \,\frac{s-\mu^2}{\bar s-\mu^2}\,\frac{B^{c+}_{ab}(\bar s)- B^{c-}_{ab}(\bar s)}{\bar s- s}
\nonumber\\
&+&\frac{1}{2\,i} \int_{C_A} \frac{d \bar s }{\pi} \,\frac{s-\mu^2}{\bar s-\mu^2}\,\frac{B^{A+}_{ab}(\bar s)- B^{A-}_{ab}(\bar s)}{\bar s- s} \,,
\label{res-B-disp}
\end{eqnarray}
can be assumed. In the next step we compute the discontinuity along the contours $C^+_c$ with 
\begin{eqnarray}
&& \frac{1}{2\,i}\,\Big[
B^{c+}_{ab}(\bar s)-B^{c-}_{ab}(\bar s)\Big] =\frac{1}{2}\,
\varsigma_{ac}(\bar s)\,\Big[
U_{cb}^{c+}(\bar s)+ U_{cb}^{c-}(\bar s)\Big]
 - \frac{1}{2}\,\varsigma_{ac}(\bar s)\,\Big[
T^{c+}_{cb}(\bar s)+T^{c-}_{cb}(\bar s)\Big]
\nonumber\\
&& \qquad \qquad \qquad \qquad \qquad \!-\, \frac{1}{2}\,
\sum_{d}\,\Big[D_{ad}^{c+}(\bar s)+D_{ad}^{c-}(\bar s)\Big] \,\sum_{e}\,T_{dc}^{c-}(\bar s)\,\rho_{ce}(\bar s)\,
T_{eb}^{c+}(\bar s) 
\nonumber\\
&& \qquad \qquad \qquad \qquad \qquad \!= \,\frac{1}{2}\,
\varsigma_{ac}(\bar s)\,\Big[
U_{cb}^{c+}(\bar s)+ U_{cb}^{c-}(\bar s)\Big]\,,
\label{discontinuity-B}
\end{eqnarray}
where we observe the cancellation of most  terms in  (\ref{discontinuity-B}). This follows from the 
defining equations for $ \varsigma_{ab}(\bar s)$ in (\ref{def-varsigma}) together with symmetry of the reaction amplitude 
\begin{eqnarray}
&& U_{ab}(s) = U_{ba}(s) \!\qquad \& \qquad  \rho_{ab}(s) = \rho_{ba}(s) \,,
\label{res-symmetry}\\
\qquad \to \qquad &&  T_{ab}(s) = T_{ba}(s) \qquad \& \qquad \sum_{e}\,T^{c-}_{dc}(\bar s)\,\rho_{ce}(\bar s)\,T^{c+}_{eb}(\bar s) =
\sum_{e}\,T^{c+}_{dc}(\bar s)\, \rho_{ce}(\bar s)\,T^{c-}_{eb}(\bar s) \,. \nonumber
\end{eqnarray}
It is left to compute the discontinuity of the $B$ functions along the straight contour $C_A$, i.e. the second term in (\ref{res-B-disp}) is considered.
For any $\bar s \in C_A$ we derive
\begin{eqnarray}
&& \frac{1}{2\,i}\,\Big[
B^{A+}_{ab}(\bar s)-B^{A-}_{ab}(\bar s)\Big] =\frac{1}{2}\,\sum_c
\varsigma_{ac}(\bar s)\,\Big[
U_{cb}^{c+}(\bar s)+ U_{cb}^{c-}(\bar s)\Big]\,\Theta \big[\bar s +i\,\epsilon -\mu_c^2 \big]
\nonumber\\
&&  \qquad \qquad \qquad\,\, -\, \frac{1}{2}\,\sum_c\,\varsigma_{ac}(\bar s)\,\Big[
T^{c+}_{cb}(\bar s)+T^{c-}_{cb}(\bar s)\Big]\,\Theta \big[\bar s +i\,\epsilon -\mu_c^2 \big]
\nonumber\\
&&  \qquad \qquad \qquad\,\, -\, \frac{1}{2}\,
\sum_{c,d}\,\Big[D_{ad}^{c+}(\bar s)+D_{ad}^{c-}(\bar s)\Big] \,\sum_{e}\,T_{dc}^{c-}(\bar s)\,\rho_{ce}(\bar s)\,
T_{eb}^{c+}(\bar s) \,\Theta \big[\bar s +i\,\epsilon -\mu_c^2 \big]
\nonumber\\
&&  \qquad \qquad \qquad \,\,= \,\frac{1}{2}\,
\sum_c\varsigma_{ac}(\bar s)\,\Big[
U_{cb}^{c+}(\bar s)+ U_{cb}^{c-}(\bar s)\Big]\,\Theta \big[\bar s +i\,\epsilon -\mu_c^2 \big]\,,
\label{discontinuity-B:Acase}
\end{eqnarray}
where the cancellations in (\ref{discontinuity-B:Acase}) follow from (\ref{def-varsigma}, \ref{res-symmetry}).
We observe that owing to the 
identities (\ref{discontinuity-B}, \ref{discontinuity-B:Acase})  the two terms in (\ref{res-B-disp}) can be combined into a dispersion integral written in terms of the  partially overlapping contours $C_c$. It holds
\begin{eqnarray}
&&  B_{ab}(s) =\sum_c\int_{C_c} \frac{d \bar s }{\pi} \,\frac{s-\mu^2}{\bar s-\mu^2}\,\frac{\varsigma_{ac}(\bar s)}{\bar s- s}\,
U_{cb}(\bar s)\,,
\nonumber\\ 
&& \quad {\rm with}\qquad U_{ab}(s) = U_{ab}^{c+}( s)= U_{ab}^{c-}( s) \qquad {\rm for} \qquad s \in C_c \qquad {\rm given\; any} \quad c\,,
\label{final-res-B-disp}
\end{eqnarray}
where the crucial identity in the last line of (\ref{final-res-B-disp}) 
is a consequence of the properly deformed contour lines as illustrated in Fig. \ref{fig:3}.

With this we arrive at the anticipated representation of the scattering amplitude in terms of the spectral density $ \varsigma_{ab}(\bar s)$. It holds
\begin{eqnarray}
&& T_{ab}(s) =  U_{ab}(s) - \sum_c D^{-1}_{ac}(s) \,B_{cb}(s) 
\nonumber\\
&& \qquad \quad\! =\, U_{ab}(s) - \sum_{c,d} D^{-1}_{ad}(s) \,
\int_{C_c} \frac{d \bar s }{ \pi} \,\frac{s-\mu^2}{\bar s-\mu^2}\,\frac{\varsigma_{dc}(\bar s)\,}{\bar s- s} \,
U_{cb}(\bar s)\,,
\label{res-T-varsigma}
\end{eqnarray}
The functions $D_{ab}(s)$ were already expressed in terms of $\varsigma_{ab}(\bar s)$ in (\ref{def-varsigma}).

We are one step before a more practical rewrite of the defining request (\ref{def-varsigma}) and in particular checking the 
consistency of the construction. After all we had to use both equations in the first line of   (\ref{def-varsigma}) as to arrive at (\ref{res-T-varsigma}). 
Inserting our result  (\ref{res-T-varsigma}) into (\ref{def-varsigma}) we will obtain two distinct equations, for which we have to show their equivalence. 
We derive
\begin{eqnarray}
&& \varsigma_{ab}(s) 
=- \sum_d \, U_{ad}(s) \,\rho_{db}(s)  
\nonumber\\
&& \qquad \quad\;\;-\, \sum_{c,d} \int_{C_c} \frac{d \bar s}{\pi} \,\frac{s-\mu^2}{\bar s-\mu^2}\,
\frac{\varsigma_{ac}(\bar s)}{\bar s-s_\pm}\,\Big[ U_{cd}(s) - U_{cd}(\bar s)\Big] \,\rho_{db}( s)\,,
\label{res-varsigma}
\end{eqnarray}
where $s$ is on the contour $C_b$ strictly. With $s_\pm$  we introduce values of $s$ slightly above and below 
the contour $C_b$, i.e. it holds $| s-s_\pm | < \epsilon' $, where $\epsilon'$ is chosen smaller than the minimal distance of any of the horizontal lines in Fig. \ref{fig:3}. 
The two choices correspond to the two identities in the first line of   (\ref{def-varsigma}). 
Since the numerator in (\ref{res-varsigma}) strictly vanishes at $\bar s=s$ both choices lead to identical results. 

While with (\ref{res-varsigma}) we arrive at a mathematically well defined linear integral equation, it remains to construct 
a numerical solution to it. This is not quite straight forward and will require further developments. 
In the following we will analyze the linear system (\ref{res-varsigma}) in more detail and eventually establish a framework that 
can be used to numerically solve it on a computer. The key issue is to systematically perform the 
limit $\epsilon \to 0$ in the system of complex contours. 

\section{From complex contours to real contours}

We consider first the $D$ function, for which its definition is recalled with
\begin{eqnarray}
 && D_{ab}(s) =\delta_{ab} + \int_{C_b} \frac{d \bar s }{\pi} \,\frac{s-\mu^2}{\bar s-\mu^2}\,
\frac{\varsigma_{ab}(\bar s)}{\bar s-s }\,,
\label{recall-D}
\end{eqnarray}
in terms of the spectral weight $\varsigma_{ab}(s)$. We consider the limit $\epsilon \to 0$, in which the complex contours $C_c$ all approach the real axis. 
If we are interested in values of $s$ only that are below or above all right-hand cut lines, we may simplify the 
integral into Riemann sums on the real axis. This is achieved as follows.

We first note that we may consider $\varsigma_{ab}(s)$ to be an analytic function in 
$s$, with various branch cuts. This follows from the integral representation (\ref{res-varsigma}). More precisely, if the linear system has a solution $\varsigma_{ab}(s)$
with $s \in C_b$ then the equation  (\ref{res-varsigma}) can be used to analytically continue $\varsigma_{ab}(s)$ away from the contour line $C_b$. The branch cuts are readily  
identified. First it carries the branch cuts of the phase-space function $\rho_{cc}(s)$ that is on the real axis strictly in our convention. Second, the cut-lines of the generalized potential $U_{cb}(s)$ for any $c$ are inherited. The important observation is the absence of any right-hand cut-lines.

According to the cut lines summarized in Fig. \ref{fig:3} there are two critical points, $ \mu_b$ and $\hat \mu_b $, associated with a normal threshold point at $m_b+M_b$. While $\mu_b$ denotes the smallest anomalous threshold opening of the generalized potential $U_{ab}(s)$ with arbitary $a$, the return point $\hat \mu_b$, specifies the point at which the left-hand contour line circles around the point $s= (m_b + M_b)^2$ in Fig. \ref{fig:3} and returns back. With this in mind we introduce
\begin{eqnarray}
 && D_{ab}(s) =\delta_{ab} + \int^\infty_{(m_b +M_b)^2} \frac{d \bar s }{\pi} \,\frac{s-\mu^2}{\bar s-\mu^2}\,
\frac{\hat \varsigma_{ab}(\bar s)}{\bar s-s } + 
\int^{\hat  \mu^2_b}_{\mu_b^2} \frac{d \bar s }{\pi} \,\frac{s-\mu^2}{\bar s-\mu^2}\,
\frac{\Delta \varsigma_{ab}(\bar s)}{\bar s-s } \,,
\nonumber\\ \nonumber\\
&& \;{\rm with } \qquad  \hat \varsigma_{ab} (\bar s) = \varsigma^+_{ab}(\bar s)\,\Theta \big[\hat  \mu^2_b - \bar s\big]\,\Theta \big[ \bar s - (m_b + M_b)^2 \big]
 + \varsigma^-_{ab}(\bar s)\,\Theta \big[\bar s -\hat \mu^2_b \big] \,,
\nonumber\\
&& \;\qquad  \qquad \Delta \varsigma_{ab}(\bar s) = \varsigma^-_{ab}(\bar s) - \varsigma^+_{ab}(\bar s)\,,
\label{rewrite-D}
\end{eqnarray}
where the integrals over $\bar s$ in (\ref{rewrite-D}) are on the real axis strictly. In (\ref{rewrite-D}) we apply the useful notation 
\begin{eqnarray}
&& \varsigma^\pm_{ab}(\bar s ) = \varsigma_{ab}(s_\pm ) \qquad {\rm with} \qquad \Im \bar s= 0 \qquad {\rm and} \qquad  s_\pm \in C_b 
\nonumber\\
&& \qquad {\rm and} \qquad \Im (s_+-s_-)  > 0 \qquad {\rm and} \qquad \Re s_\pm =\bar s\,.
\label{def-varsigma-pm}
\end{eqnarray}
which defines $ \varsigma^+_{ab}(\bar s )$ for $\bar s < (m_b + M_b)^2$ initially only, but is naturally extended up to the 
point at $\bar s = \hat \mu^2_b$. Here we assume the availability of the analytic continuation of the function $\varsigma_{ab}(s)$ from the nominal 
threshold value at $s =(m_b + M_b)^2$ up to the return point of the left-hand cut at $s = \hat \mu_b^2$ in Fig. \ref{fig:3}. The particular location of $\hat \mu_b > m_b + M_b$ 
is irrelevant. While the spectral weights $\hat \varsigma_{ab} (\bar s)$ and $\Delta \varsigma_{ab} (\bar s)$ depend on it, 
by construction the $D$ function does not. Given (\ref{def-varsigma-pm}) the limit $\epsilon \to 0$ in the right- and left-hand contours of Fig. \ref{fig:3} 
can be applied without changing the form of (\ref{rewrite-D}). In this limit the contributions of any vertical parts of the contour lines vanish. 
Such terms are already omitted altogether in (\ref{rewrite-D}). Note that modulo those vertical lines  (\ref{rewrite-D}) is nothing
but a regrouping of the various contour contributions in (\ref{recall-D}).

\begin{figure}[t]
\center{
\includegraphics[keepaspectratio,width=0.95\textwidth]{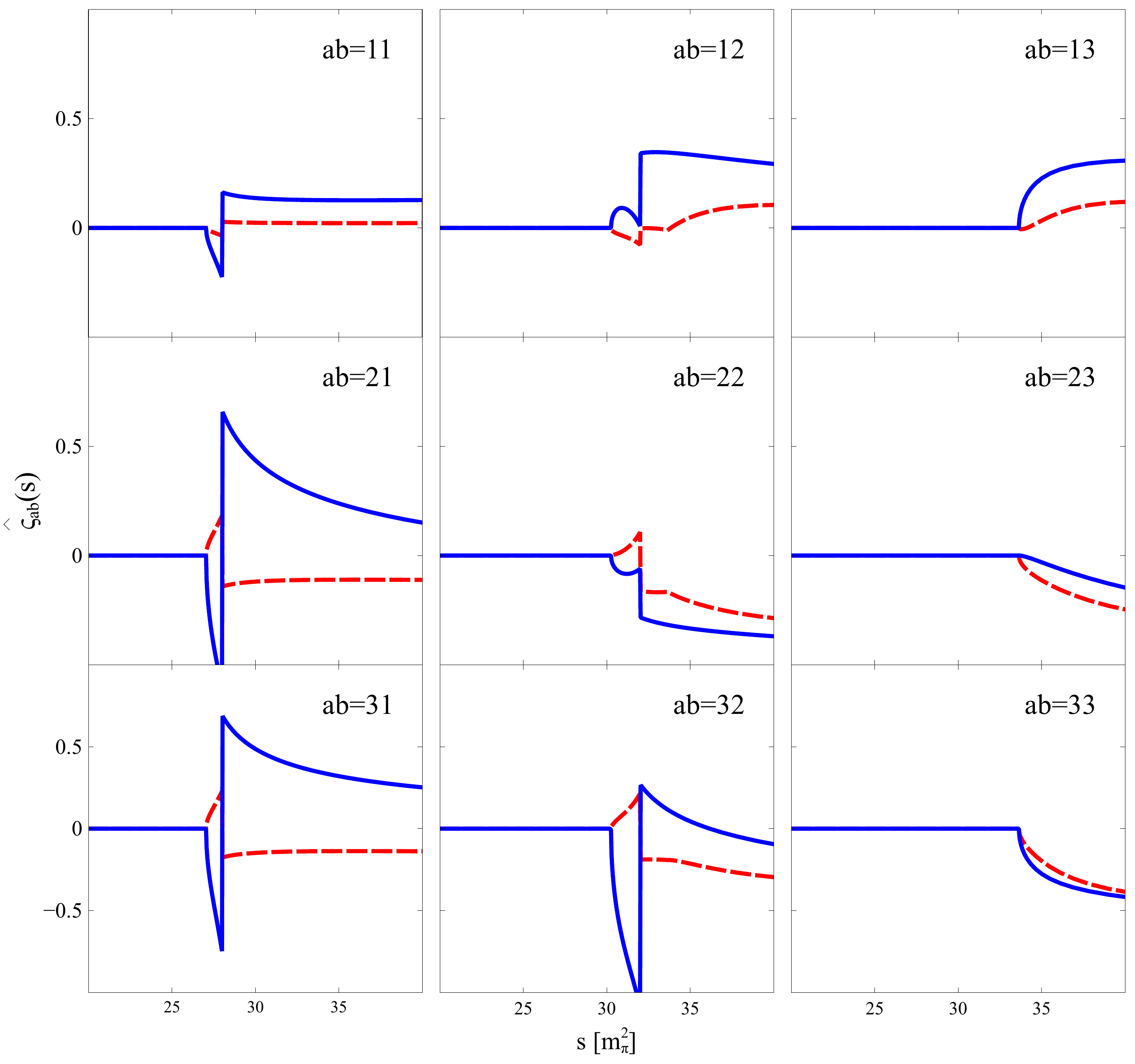} }
\caption{\label{fig:10} The function $\hat \varsigma_{ab}(s)$ of (\ref{rewrite-D}, \ref{res-varsigmahat-final}) in our schematic model (\ref{def-model}). Real and imaginary parts are shown with solid blue and dashed red lines respectively.  }
\end{figure}

We illustrate the generic form of the spectral weight $\hat \varsigma_{ab}(s)$ in our model (\ref{def-model}). As shown in Fig. \ref{fig:10} the complex functions are non-zero for $s > (m_b + M_b )^2$ only. It is important to note that the latter do depend on the choice of the return point $\hat \mu_b$. The functions $\hat \varsigma_{ab}(s)$ are piecewise continuous with the only discontinuous behavior at the return point $s =\hat \mu^2_b$.

The result (\ref{rewrite-D}) is useful since in the limit $\epsilon \to 0$ the anomalous spectral weight $\Delta \varsigma_{ac}(\bar s)$ can be linked back to 
the $D$ function as follows. A direct application of (\ref{res-varsigma}) leads to
\begin{eqnarray}
&& \Delta \varsigma_{ab}(\bar s) =  -\,\sum_{c > b,d} D_{ac}(\bar s+i\,\epsilon')\,\big[U^{-d}_{cd}(\bar s) -U^{+d}_{cd}(\bar s) \big]\,\Theta \big[ (\mu^A_{cd})^2 < \bar s < \hat \mu^2_b\big]\,\rho_{db}(\bar s)\,,
\nonumber\\
&& U^{\pm b}_{cd}(\bar s ) = U_{cd}(\bar s_\pm ) \qquad  \qquad {\rm with} \qquad \Im \bar s= 0 \qquad {\rm and} \qquad \bar s_\pm \in C_b
\nonumber\\
&& \qquad {\rm and} \qquad \Im (\bar s_+-\bar s_-)  > 0 \qquad {\rm and} \qquad \Re \bar s_\pm =\bar s\,,
 \label{res-varsigmaA}
\end{eqnarray}
where we used the crucial property that the generalized potential 
$U_{ab}(s)$ may develop an anomalous threshold behavior at the lower of the two nominal thresholds at $ s=(m_a+M_a)^2$  or $s=(m_b+M_b)^2$. Given our strict channel ordering the sum in  (\ref{res-varsigmaA}) over the channel index $c$ is restricted to the case $c > b$.

There is a subtle point as to where to evaluate the $D_{ac}(\bar s_- -i\,\epsilon')$ function in 
the first line of (\ref{res-varsigmaA}). Since  $D_{ac}(s)$ has a branch cut along $C_c$ and  
$ \bar s_- \in C_c$ for $\bar s > \mu_c ^2$ it is necessary to specify whether we should evaluate the function below or above the cut. Our prescription follows unambiguously if we slightly deform  the contours $C_c$ in (\ref{res-varsigma}). In Fig. \ref{fig:3} the solid line passing through $\bar s_-$
is deformed a bit towards, but still avoiding, the dashed lines above. With this it is manifest 
from (\ref{res-varsigma}) that $\varsigma_{ab}(s)$ is analytic along the horizontal line through $\bar s_-$. This should be so since $\varsigma_{ab}(s)$ is analytic along all contours $C_c$ by construction. In turn our prescription $D_{ac}(\bar s_- -i\,\epsilon')$ is justified. Note that for the 
term $D_{ac}(\bar s_+)$ no further specification is needed simply because $\bar s_+ \notin C_c$. Here it always holds $D_{ac}(\bar s_+) = D_{ac}(\bar s + i\,\epsilon')$.
We ask the reader to carefully discriminate the objects $ U^{c\pm}_{ab}( s)$ with $s \in C_c$ as introduced in (\ref{def-Apm}) from the 
newly introduced object $ U^{\pm c}_{ab}(\bar s)$  with $\bar s$ defined on the real axis only in (\ref{res-varsigmaA}).

In the following we will show that given the function $\hat \varsigma_{ab}(\bar s)$ only the $D_{ab}(s)$ function can be computed unambiguously. 
Note that this requires the solution of a linear integral equation since $\Delta \varsigma_{ab}(\bar s)$ requires the knowledge of the $D_{ab}(s)$ function.  
In order to solve this system it is useful to introduce some notation
\begin{eqnarray}
 && \rho^L_{ab}( \bar s) = \sum_{c<a}\big[U^{+c}_{ac}(\bar s) -U^{-c}_{ac}(\bar s) \big]\,\rho_{cb}(\bar s)\, \Theta \big[ (\mu^A_{ac})^2 < \bar s < \hat \mu^2_b\big]
\qquad \quad {\rm for}\qquad \quad a > b\,,
 \nonumber\\
 &&  \rho^L_{ab}( \bar s)  = 0 \qquad {\rm for}\qquad a \leq b  \,,
 \label{def-rhoL}
\end{eqnarray}
where $\bar s$ is on the real axis. Note that due to the particular form 
of the anomalous box term $U^{\rm box}_{ab}(s)$ in (\ref{def-Ubox}) there is no contribution from the latter to $\rho^L_{ab}( \bar s) $.
Our result  (\ref{rewrite-D}, \ref{res-varsigmaA}) is expressed  as follows
\begin{eqnarray}
 && D_{ab}(s) =\delta_{ab} + \int^\infty_{(m_b +M_b)^2} \frac{d \bar s }{\pi} \,\frac{s-\mu^2}{\bar s-\mu^2}\,
\frac{\hat \varsigma_{ab}(\bar s)}{\bar s-s } + \sum_c
\int \frac{d \bar s }{\pi} \,\frac{s-\mu^2}{\bar s-\mu^2}\,
\frac{D_{ac}(\bar s+i\,\epsilon')\,\rho^L_{cb}(\bar s)}{\bar s-s } \,,
\label{rewrite-D-second}
\end{eqnarray}
where all integrals over $\bar s$ are on the real axis. The bounds of the integrals are provided by the properties of $ \rho^L_{cb}(\bar s)$ as summarized in (\ref{def-rhoL}).

It remains to express $D_{ab}(s)$ in terms of $\hat \varsigma_{ab}(s)$ and $\rho^L_{ab}(s)$. In the non-overlapping case this is readily achieved by an iteration in the index $b$. At $b = {\rm {\rm max }}$ the second term in 
(\ref{rewrite-D-second}) does not contribute as a consequence of the second condition in (\ref{def-rhoL}). In turn we can compute $D_{a b}$ at $b = {\rm max }$ for all a. 
In the next step we study $D_{ab}(s)$ at $b = {\rm max }-1$, where now the second term in (\ref{rewrite-D-second}) turns relevant. However, here only the previously 
computed $D_{ab}(s)$ at $b= {\rm max }$ are needed. This process can be iterated down to the computation of $D_{a1}(s)$. While this strategy is always leading to the correct result it is not very efficient for the following developments.

\begin{figure}[t]
\center{
\includegraphics[keepaspectratio,width=0.65\textwidth]{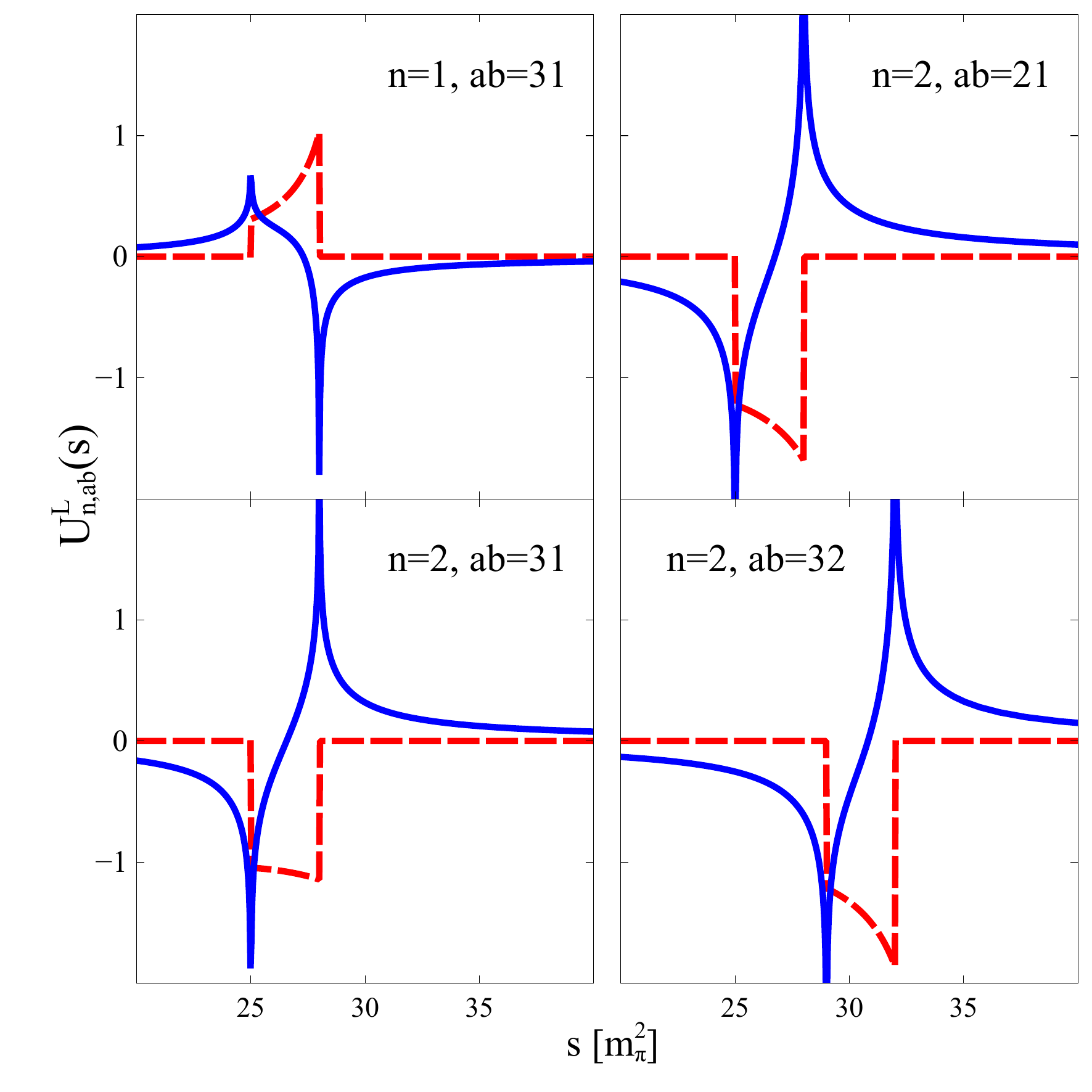} }
\caption{\label{fig:12} The non-vanishing functions $U^L_{n,ab}(s)$ of (\ref{res-ULa}) in our schematic model (\ref{def-model}). Note the relation $U^L_{3,31}(s)= U^L_{1,31}(s)$. Real and imaginary parts are shown with solid blue and dashed red lines respectively.}
\end{figure}

A more powerful framework can be readily established as follows.
We introduce a Green's function $L(x,y)$ via the condition
\begin{eqnarray}
\int dy \bigg[ \delta(x-y) - \frac{1}{\pi}\,\frac{\rho^L(y)}{x-y-i\,\epsilon'} \bigg]\,L (y,z)\, = \delta(x-z)\,,
\label{def-Lxy}
\end{eqnarray}
where we suppress the coupled-channel matrix structure for notational clarity. All objects will be written in the correct order, so that 
the matrix structure can be reconstructed  unambiguously for any identity presented below. The $i\,\epsilon'$ prescription in the definition of  
$L(x,y)$ in (\ref{def-Lxy}) is inherited from the $i\,\epsilon'$ prescription in (\ref{rewrite-D-second}). 
Given the Green's function $D_{ab}(s)$ can be expressed 
in terms of $\hat \varsigma_{ab}(s)$ with
\begin{eqnarray}
&& D_{ab}(s) = \delta_{ab} +  \int \frac{d \bar s }{\pi} \,\frac{s-\mu^2}{\bar s-\mu^2}\,
\frac{\varsigma^D_{ab}(\bar s)}{\bar s- s } \,,
\nonumber\\
&&  \varsigma^D_{ab}(s)  = \sum_c
\int d \bar s  \,\frac{s- \mu^2}{\bar s- \mu^2} \,\Big( \Theta \big[\bar s- (m_c+ M_c)^2\big]\,\hat \varsigma_{ac} (\bar s) + \rho^L_{ac}(\bar s)\Big)\,L_{cb}(\bar s, s)\,.
\label{res-D-from-L}
\end{eqnarray}
The representation (\ref{res-D-from-L}) does not look very promising for numerical simulations since the Green's function is a highly singular object. 
However,  a closed form  can be derived in terms of six analytic matrix functions $u_n^L(s)$ and $U_n^L(s)$ with $n=1,2,3$. 
The latter are determined by appropriate integrals involving the anomalous spectral weight $\rho^L(s)$ only.
After some algebra we established the following form for the Green's function
\begin{eqnarray}
 L(x,y) = \delta(x-y) +\frac{1}{\pi}\,\frac{\rho^L(y)}{x-y-i\,\epsilon'} + \sum_{n=1}^3\,u_n^L(x)\,\frac{U_n^L(x)-U_n^L(y)}{x-y}\,\frac{1}{\pi}\, \rho^L(y)\,,
\end{eqnarray}
where
\begin{figure}[t]
\center{
\includegraphics[keepaspectratio,width=0.95\textwidth]{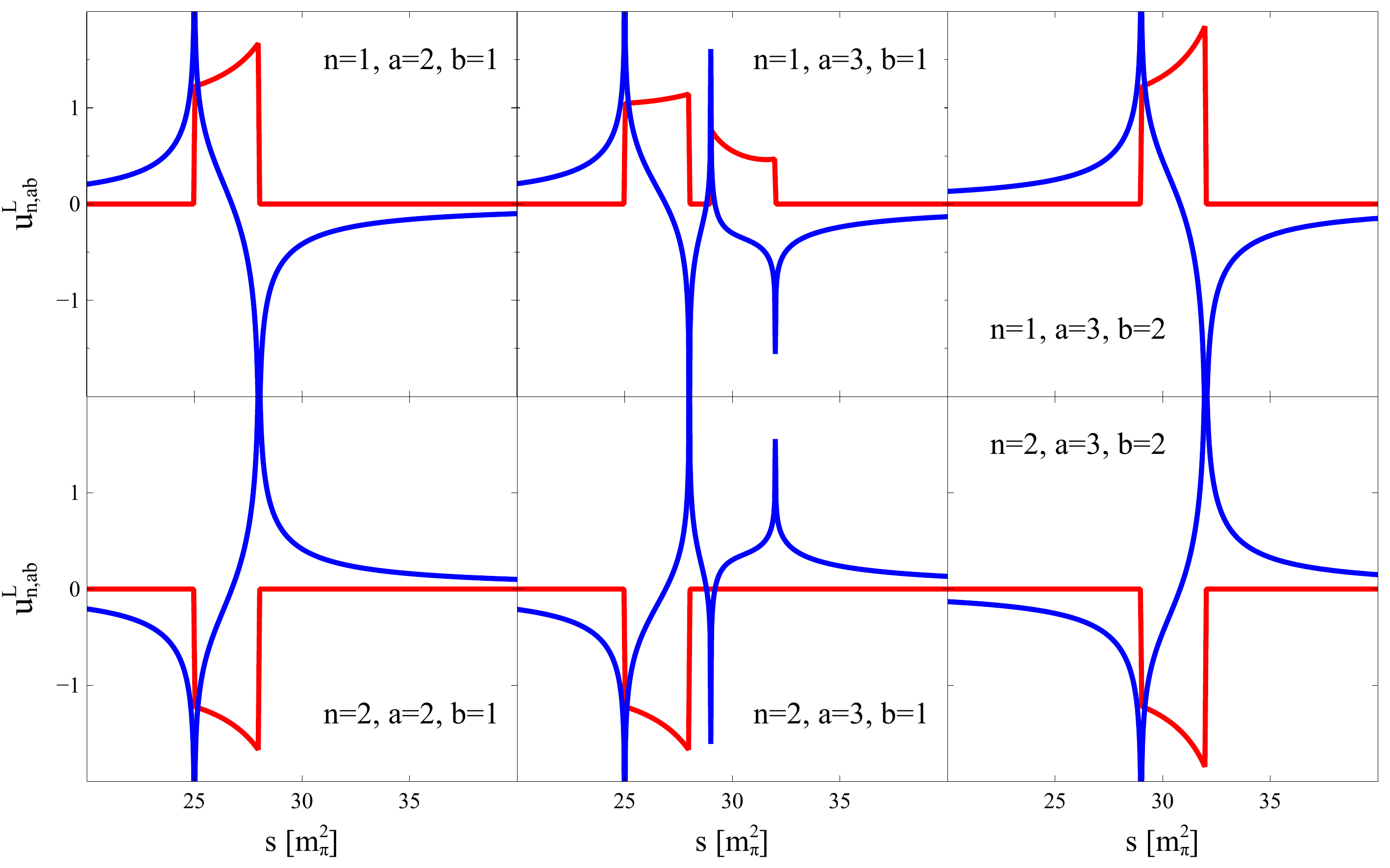} }
\caption{\label{fig:11} The non-vanishing functions $u^L_{n,ab}(s)$ of (\ref{res-ULa}) for $n=1,2$ in our schematic model (\ref{def-model}). Real and imaginary parts are shown with solid blue and dashed red lines respectively. }
\end{figure}
\allowdisplaybreaks[1]
\begin{eqnarray}
&& u_1^L(x) = g^L(x) -\int\,\frac{d z}{\pi }\,\frac{\rho^L(z)\,g^L(z)}{z-x+i\,\epsilon'} \,, \qquad \quad 
u_2^L(x) =h^L(x) -\int\,\frac{d z}{\pi }\,\frac{\rho^L(z)\,h^L(z)}{z-x+i\,\epsilon'} \,,
\nonumber\\
&&  U^L(x) = \int\,\frac{d z}{\pi }\,\frac{\rho^L(z)}{z-x -i\,\epsilon'}\qquad \qquad \qquad \qquad \quad \quad \!\! 
u_3^L(x) = -1\,,  
\nonumber\\
&& \quad {\rm with}\qquad g^L(x) = U^L(x) -\int\,\frac{d z}{\pi }\,\Big[U^L(z)-U^L(x)\Big]\,\frac{\rho^L(z )}{z- x}\, g^{L}(z) \,,
\nonumber\\
&& \qquad \qquad \quad \!h^L(x) = 1 -\int\,\frac{d z}{\pi }\,\Big[U^L(z)-U^L(x)\Big]\,\frac{\rho^L(z )}{z- x}\, h^{L}(z)\,, 
\label{res-uLa}\\ \nonumber\\
&& U_1^L(x) = \int \frac{d z}{\pi }\,\frac{\rho^L(z)}{z-x}\,\Big[  U^L(z )\,\Delta U_1^L(z,x)+ \Delta U_2^L(z,x)\Big]\,,
\nonumber\\
&& U_2^L(x) =-U^L(x)
-\int \frac{d z}{\pi }\,\frac{U^L(z)\,\rho^L(z)}{z-x}\,\Big[U^L(z )\,\Delta U_1^L(z,x)+ \Delta U_2^L(z,x) \Big]\,,
\nonumber\\
&& U_3^L(x) = \int \frac{ d z}{\pi}\,\frac{\rho^L(z)\,g^L(z)}{z-x}\,
\Delta U_1^L(z,x) + \, \int \frac{ d z}{\pi}\,\frac{\rho^L(z)\,h^L(z)}{z-x}\,
\Delta  U_2^L(z,x)\,,
\nonumber\\ \nonumber\\
&& \quad {\rm with}\qquad \Delta U^L_n (z,x) = U^L_n (z)- U^L_n (x)\,.
\label{res-ULa}
\end{eqnarray}
We illustrate the general form of the complex objects $u^L_n(s)$ and $U^L_n(s)$ at hand of our schematic model (\ref{def-model}). In Fig. \ref{fig:11} and Fig. \ref{fig:12} all elements are shown.

An important property of our result (\ref{res-D-from-L}) is that the imaginary part of the spectral weight $\varsigma^D_{ab}(\bar s)$ does not vanish in the presence of anomalous threshold effects. 
This follows from the results (\ref{res-D-from-L}-\ref{res-ULa}). In turn the functions $D_{ab}(s)$ do not satisfy the Schwarz reflection principle  with
\begin{eqnarray}
 D^*_{ab}(s) \neq D_{ab}(s^*) \qquad \leftrightarrow \qquad  \Im\hat \varsigma_{ab}(\bar s) \neq - \sum_c \Im \big[ D_{ac}(\bar s + i\,\epsilon) \,\rho^L_{cb}(\bar s) \big] \,,
 \label{D-Schwarz}
\end{eqnarray}
even in the limit with $\epsilon \to 0$. We emphasize that this is unavoidable in our formulation. 
Nevertheless we expect that our final reaction amplitudes $T^*_{ab}(s) = T_{ab}(s^*)$ will satisfy the Schwarz reflection principle, which plays a crucial role in the derivation of the coupled-channel unitarity condition. Note that (\ref{D-Schwarz}) should not be taken as a surprise since we have already discussed that the anomalous box term $U^{\rm box}_{ab}( \bar s)$ is at odds with (\ref{def-Schwarz}). The relation (\ref{D-Schwarz}) can be confirmed explicitly upon an expansion of 
the $D$-function in powers of the generalized potential. At second order there is a term that confirms (\ref{D-Schwarz}).

We continue with the $B$ function, for which its representation 
\begin{eqnarray}
 && B_{ab}(s) = \sum_c \int_{C_c} \frac{d \bar s }{\pi} \,\frac{s-\mu^2}{\bar s-\mu^2}\,
\frac{\varsigma_{ac}(\bar s)}{\bar s-s }\,U_{cb}(\bar s)\,,
\label{recall-B}
\end{eqnarray}
in terms of the spectral weight $\varsigma_{ac}(s)$ and the generalized potential $U_{ac}(s)$ is recalled. 
Again we perform the limit $\epsilon \to 0$, in which the complex contours $C_c$ all approach the real axis.

Following the decomposition of the contour lines $C_c$ as introduced in our study of the $D$ function in (\ref{rewrite-D}) we readily derive the corresponding rewrite 
\begin{eqnarray}
 && B_{ab}(s) = \sum_c \int^\infty_{(m_c +M_c)^2} \frac{d \bar s }{\pi} \,\frac{s-\mu^2}{\bar s-\mu^2}\,
\frac{\hat \beta^{\,c}_{ab}(\bar s)}{\bar s-s } + \sum_c
\int^{\hat  \mu^2_c}_{\mu_c^2} \frac{d \bar s }{\pi} \,\frac{s-\mu^2}{\bar s-\mu^2}\,
\frac{\Delta \beta^{\,c}_{ab}(\bar s)}{\bar s-s } \,,
\nonumber\\ \nonumber\\
&& \;{\rm with } \qquad   \hat \beta^{\,c}_{ab} (\bar s) = \varsigma^+_{ac}(\bar s)\,U^{+c}_{cb}(\bar s)\Theta (\hat  \mu^2_c - \bar s)
 + \varsigma^-_{ac}(\bar s)\,U^{-c}_{cb}(\bar s)\,\Theta (\bar s -\hat \mu^2_c ) \,,
\nonumber\\
&& \;\qquad  \qquad \Delta \beta^{\,c}_{ab}(\bar s) = \varsigma^-_{ac}(\bar s)\,U^{-c}_{cb}(\bar s) - \varsigma^+_{ac}(\bar s)\,U^{+c}_{cb}(\bar s)\,,
\label{rewrite-B}
\end{eqnarray}
where the integrals over $\bar s$ in (\ref{rewrite-B}) are on the real axis strictly. Again modulo contributions in (\ref{recall-B}) from vertical parts the  
contours $C_c$ both representations (\ref{recall-B}) and (\ref{rewrite-B}) are identical at any finite $\epsilon$. In (\ref{rewrite-B}) we apply the 
convenient notation $\varsigma^\pm_{ac}(\bar s)$ and $U^{\pm c}_{cb}(\bar s)$ introduced already in (\ref{def-varsigma-pm}, \ref{res-varsigmaA}).

In the following we will express the spectral weights $\hat \beta^{\,c}_{ab} (\bar s)$ and $\Delta \beta^{\,c}_{ab} (\bar s)$ in terms of 
$ \hat \varsigma_{ab} (\bar s)$ and the $\Delta \varsigma_{ab} (\bar s)$. For this we will have to consider the limit $\epsilon \to 0$ again, in which 
we find
\begin{eqnarray}
&&\hat \beta^{\,c}_{ab} (\bar s) = \hat  \varsigma_{ac}(\bar s)\,U_{cb}(\bar s) \qquad {\rm for }\qquad \bar s > (m_c+M_c)^2\,,
\label{res-hatbeta}
\end{eqnarray}
where the generalized potential $U_{cb}(\bar s)$ is evaluated on the real axis strictly. It is important to realize that $U_{cb}(\bar s)$ 
is needed in between the upper and lower anomalous  cut lines (dashed lines in Fig. \ref{fig:2}) even after the limit $\epsilon \to 0$ has been performed. 
For the anomalous spectral weight  $\Delta \beta^{\,c}_{ab}(\bar s) $ we derive three distinct contributions
\begin{eqnarray}
&& \Delta \beta^{\,c}_{ab}(\bar s) =  
   \Delta \varsigma^{}_{ac}(\bar s)\,U^{-c}_{cb}(\bar s)  
+  \varsigma^+_{ac}(\bar s) \Big[U^{-c}_{cb}(\bar s) - U^{+c}_{cb}(\bar s) \Big] 
\qquad \qquad \quad {\rm for} \qquad  b> c  \,,
\nonumber\\
&& \Delta \beta^{\,c}_{ab}(\bar s) =    \Delta \varsigma^{}_{ac}(\bar s)\,U_{cb}(\bar s)     \qquad \qquad \qquad \qquad \qquad \qquad \qquad \qquad \quad \, {\rm for} \qquad  b= c \,,
\nonumber\\
&& \Delta \beta^{\,c}_{ab}(\bar s) = \Delta \varsigma^{}_{ac}(\bar s)\,U_{cb}(\bar s) 
\nonumber\\
&& \qquad \quad \; \;\, +\,   \Theta \big[\bar s -\mu_c^2 \big]\,\Theta \big[\hat \mu_b^2-\bar s\big]\, \Delta \varsigma_{ac} (\bar s)
 \Big[ U^{-b}_{cb}(\bar s) - U^{+b}_{cb}(\bar s)\Big]
\qquad \;\; \,{\rm for} \qquad  b< c \,,
 \label{res-deltabeta}
\end{eqnarray}
which contribute depending on the various cases $b > c$, $b  <c$ or $b = c$. It is important to note that in (\ref{res-deltabeta}) the contribution of the box term (\ref{def-Ubox}) 
requires particular attention. Its contribution to $U_{cd}(\bar s)$ is to be taken slightly below the double cut structure of (\ref{def-Ubox}). In the following we will promote this to our convention. Whenever we use $U_{cd}(s)$ with $s$ on the real axis the above prescripton is implied. 
We express our results  (\ref{rewrite-B}, \ref{res-hatbeta}, \ref{res-deltabeta}) in the notation (\ref{def-rhoL}) as follows
\begin{eqnarray}
 && B_{ab}(s) = \sum_c \int_{(m_c+ M_c)^2}^\infty \frac{d \bar s }{\pi} \,\frac{s-\mu^2}{\bar s-\mu^2}\,
\frac{ \hat \varsigma_{ac}(\bar s)}{\bar s-s }\,U_{cb}(\bar s) 
\nonumber\\
&& \qquad +\, \sum_{c\geq b,d}
\int \frac{d \bar s }{\pi} \,\frac{s-\mu^2}{\bar s-\mu^2}\,
\frac{D_{ad}(\bar s+i\,\epsilon')\,\rho^L_{dc}(\bar s)}{\bar s-s } \,U_{cb}(\bar s) 
\nonumber\\
&& \qquad +\, \sum_{c<b,d}
\int \frac{d \bar s }{\pi} \,\frac{s-\mu^2}{\bar s-\mu^2}\,
\frac{D_{ad}(\bar s+i\,\epsilon')\,\rho^L_{dc}(\bar s)}{\bar s-s } \,U^{-c}_{cb}(\bar s) 
\nonumber\\
&& \qquad +\, \sum_{c< b} \int_{\mu_c^2}^{\hat \mu_c^2} \frac{d \bar s }{\pi} \,\frac{s-\mu^2}{\bar s-\mu^2}\,
\frac{\varsigma^+_{ac}(\bar s) }{\bar s-s }\,\Big[U^{-c}_{cb}(\bar s) - U^{+c}_{cb}(\bar s) \Big]
\nonumber\\
&& \qquad+\, \sum_{d> c > b}
\int_{\mu_c^2}^{\hat \mu_b^2} \frac{d \bar s }{\pi} \,\frac{s-\mu^2}{\bar s-\mu^2}\,
\frac{D_{ad}(\bar s+i\,\epsilon)\,\rho^L_{dc}(\bar s)}{\bar s-s } \,\Big[U^{-b}_{cb}(\bar s) - U^{+b}_{cb}(\bar s) \Big] \,,
\label{rewrite-B-second}
\end{eqnarray}
where all integrals over $\bar s$ are on the real axis. The bounds of the integrals are provided directly or by the properties of  $\rho^L_{cb}(\bar s)$ as summarized in (\ref{def-rhoL}).

We point at an important subtlety. While in the previous section we managed to express $D_{ab}(s)$ in terms of $ \hat \varsigma_{ab}(s)$ at $s > (m_b + M_b)^2$, this is  not 
possible for $B_{ab}(s)$. The term in (\ref{rewrite-B-second}) involving $\varsigma^+_{ac}(s)$ is required at $ \mu_c^2 <s < \hat \mu_c^2$ partially outside the domain 
where $\hat \varsigma_{ac}(s)$  was introduced in (\ref{rewrite-D}). 
It is instrumental to introduce a function $\varsigma^{\circ}_{ac}(s)$ defined on that interval, that coincides with $\hat \varsigma_{ac}(s)$ in the region where their defining domains overlap.  We do so as follows
\begin{eqnarray}
&& \varsigma^+_{ac}(s) =  \varsigma^{\circ}_{ac}(s) - \sum_e \Big[ U^{-a}_{ae}(s) -  U^{+a}_{ae}(s) \Big]\rho_{ec}( s) 
\nonumber\\
&& \qquad \;\; -\, 
\sum_{d < c} \int \frac{d\,\bar s}{\pi}\,\frac{s-\mu^2}{\bar s-\mu^2}\,\frac{\hat \varsigma_{ad}(\bar s)}{\bar s-s -i\,\epsilon_{de}'}\,\Big[ U^{-d}_{de}(s) -  U^{+d}_{de}(s) \Big]\rho_{ec}( s)
\nonumber\\
&& \qquad\;\; = \, \varsigma^\circ_{ac}(s) -  \sum_{d < c} D^{\,e}_{ad}(s )\,
\Big[ U^{-d}_{de}(s) -  U^{+d}_{de}(s) \Big]\rho_{ec}( s)
\,,
\label{def-varsigma-circ}
\end{eqnarray}
with $\mu_c^2< s < \hat \mu_c^2$ on the real axis strictly and 
\begin{eqnarray}
 && D^{\,e}_{ad}(s) =\delta_{ad} + \int^\infty_{(m_d +M_d)^2} \frac{d \bar s }{\pi} \,\frac{s-\mu^2}{\bar s-\mu^2}\,
 \left( 
\frac{\varsigma^-_{ad}(\bar s)\Theta[ \bar s -\hat \mu_d^2 ]}{\bar s-s -i\,\epsilon'} 
+ \frac{\varsigma^+_{ad}(\bar s)\Theta[\hat \mu_d^2 -\bar s ]}{\bar s-s -i\,\epsilon_{de}'}\right) \,
\nonumber\\
&& \qquad \quad \, =\,\delta_{ad} + \int^\infty_{(m_d +M_d)^2} \frac{d \bar s }{\pi} \,\frac{s-\mu^2}{\bar s-\mu^2}\,
  \frac{\hat \varsigma_{ad}(\bar s)}{\bar s-s -i\,\epsilon_{de}'}\,,
\nonumber\\
&& \qquad  {\rm with } \qquad \qquad
 \epsilon_{de}' = 
\left\{
\begin{array}{lll}
-\,\epsilon' & \quad {\rm for } &  \mu_e^2 < s < \hat \mu_d^2  \\
+\,\epsilon' & \quad {\rm else}
\end{array}
\right. \;\,.
\label{rewrite-D-third}
\end{eqnarray}
Note that while the integral over $\bar s$ in (\ref{def-varsigma-circ}) is on the complex contour $C_d$ it is on the real axis in (\ref{rewrite-D-third}) strictly. 

It may be useful to point out that for a non-overlapping case system  $\hat \mu_n <  \mu_{n+1}$ for all $n$, it follows $\varsigma^\circ_{ac}(s) = \varsigma^+_{ac}(s)$.
We recall that 
$\hat \varsigma_{ac}(s)$ was introduced in $ (m_c + M_c)^2 < s < \hat \mu_c^2$ to be the analytic continuation of $\varsigma_{ac}^+(s)$. In turn it holds:
\begin{eqnarray}
&&\Theta \big[ s- \hat \mu_c^2\big]\, \hat \varsigma_{ac}(s)
+\Theta \big[ \hat \mu_c^2 - s \big]\, \varsigma^\circ_{ac}(s)
=  \Theta \big[ s- (m_c+M_c)^2\big]\, \hat \varsigma_{ac}(s)
\nonumber\\
&& \qquad \qquad \qquad \qquad \qquad \qquad \qquad \quad \;+\,\Theta \big[(m_c+M_c)^2 - s \big]\, \varsigma^\circ_{ac}(s)  \,.
\label{varsigma-circ-property}
\end{eqnarray}
With this we arrive at the central result of this section
\begin{eqnarray}
 && B_{ab}(s) = \sum_c \int_{(m_c+ M_c)^2}^\infty \frac{d \bar s }{\pi} \,\frac{s-\mu^2}{\bar s-\mu^2}\,
\frac{ \hat \varsigma_{ac}(\bar s)}{\bar s-s }\,U_{cb}(\bar s) 
\nonumber\\
&& \qquad +\, \sum_{c\geq b,d}
\int \frac{d \bar s }{\pi} \,\frac{s-\mu^2}{\bar s-\mu^2}\,
\frac{D_{ad}(\bar s+i\,\epsilon')\,\rho^L_{dc}(\bar s)}{\bar s-s } \,U_{cb}(\bar s) 
\nonumber\\
&& \qquad +\, \sum_{c<b,d}
\int \frac{d \bar s }{\pi} \,\frac{s-\mu^2}{\bar s-\mu^2}\,
\frac{D_{ad}(\bar s+i\,\epsilon')\,\rho^L_{dc}(\bar s)}{\bar s-s } \,U^{-c}_{cb}(\bar s) 
\nonumber\\
&& \qquad +\, \sum_{c< b} \int_{\mu_c^2}^{\hat \mu_c^2} \frac{d \bar s }{\pi} \,\frac{s-\mu^2}{\bar s-\mu^2}\,
\frac{ \varsigma^\circ_{ac}(\bar s) }{\bar s-s }\,\Big[U^{-c}_{cb}(\bar s) - U^{+c}_{cb}(\bar s) \Big]
\nonumber\\
&& \qquad +\, \sum_{d< c < b}
\int_{\mu_c^2}^{\hat \mu_c^2} \frac{d \bar s }{\pi} \,\frac{s-\mu^2}{\bar s-\mu^2}\,
\frac{D^{\,c}_{ad}(\bar s)}{\bar s-s } \,\Big[U^{-d}_{dc}(\bar s) - U^{+d}_{dc}(\bar s) \Big] \,\rho^R_{cb}(\bar s) 
\nonumber\\
&& \qquad+\, \sum_{d> c> b}
\int_{\mu_c^2}^{\hat \mu_b^2} \frac{d \bar s }{\pi} \,\frac{s-\mu^2}{\bar s-\mu^2}\,
\frac{D_{ad}(\bar s+i\,\epsilon)}{\bar s-s }\,\rho^L_{dc}(\bar s) \,\Big[U^{-b}_{cb}(\bar s) - U^{+b}_{cb}(\bar s) \Big]\,,
\label{rewrite-B-second}
\end{eqnarray}
in terms of $\varsigma^\circ_{ac}(\bar s)$ and $D^{\,c}_{ad}(\bar s) $ as introduced in 
(\ref{def-varsigma-circ}, \ref{rewrite-D-third}) and 
\begin{eqnarray}
 && \rho^R_{ab}( \bar s) = \sum_{c<b}\rho_{ac}(\bar s)\,  \big[U^{+c}_{cb}(\bar s) -U^{-c}_{cb}(\bar s) \big]\,\Theta \big[ (\mu^A_{cb})^2 < \bar s < \hat \mu^2_a\big]
\qquad \quad {\rm for}\qquad \quad a < b\,,
 \nonumber\\
 &&  \rho^R_{ab}( \bar s)  = 0 \qquad {\rm for}\qquad a \geq b  \,.
 \label{def-rhoL}
\end{eqnarray}
 
It is useful to provide yet a further rewrite of the $B$-function that further 
streamlines the various prescriptions. We find
 \allowdisplaybreaks[1]
\begin{eqnarray}
 && B_{ab}(s) = \sum_c \int_{(m_c+ M_c)^2}^\infty \frac{d \bar s }{\pi} \,\frac{s-\mu^2}{\bar s-\mu^2}\,
\frac{ \hat \varsigma_{ac}(\bar s)}{\bar s-s }\,U_{cb}(\bar s) 
\nonumber\\
&& \qquad +\, \sum_{c,d}
\int \frac{d \bar s }{\pi} \,\frac{s-\mu^2}{\bar s-\mu^2}\,
\frac{D_{ad}(\bar s+i\,\epsilon')\,\rho^L_{dc}(\bar s)}{\bar s-s } \,\bar U_{cb}(\bar s) 
\nonumber\\
&& \qquad -\, \sum_{c< b} \int_{\mu_c^2}^{\hat \mu_c^2} \frac{d \bar s }{\pi} \,\frac{s-\mu^2}{\bar s-\mu^2}\,
\frac{ \varsigma^\circ_{ac}(\bar s) }{\bar s-s }\,\Delta U_{cb}(\bar s) 
\nonumber\\
&& \qquad -\, \sum_{d< c < b}
\int_{\mu_c^2}^{\hat \mu_c^2} \frac{d \bar s }{\pi} \,\frac{s-\mu^2}{\bar s-\mu^2}\,
\frac{D^{\,c}_{ad}(\bar s)}{\bar s-s } \,\Delta U_{dc}(\bar s) \,\rho^R_{cb}(\bar s) 
\nonumber\\
&& \qquad-\, \sum_{d> c> b}
\int_{\mu_c^2}^{\hat \mu_b^2} \frac{d \bar s }{\pi} \,\frac{s-\mu^2}{\bar s-\mu^2}\,
\frac{D_{ad}(\bar s+i\,\epsilon)}{\bar s-s }\,\rho^L_{dc}(\bar s) \,\Delta U_{cb}(\bar s)\,,
\label{rewrite-B-third-old}
\end{eqnarray} 
 with
\begin{eqnarray}
&& \Delta U_{ab}(\bar s) = \Theta \big[ a - b \big]\,\Big[U^{+b}_{ab}(\bar s) - U^{-b}_{ab}(\bar s) \Big]
 +\Theta \big[  b -a  \big]\,\Big[U^{+a}_{ab}(\bar s) - U^{-a}_{ab}(\bar s) \Big]
\,,
\nonumber\\
&& \bar U_{ab}(\bar s) =  \Theta \big[ a - b \big]\,U_{ab}(\bar s) 
 +\Theta \big[  b -a  \big]\,U_{ab}(\bar s -i \,\epsilon)\,, 
 \label{def-Delta-U}
\end{eqnarray}
where we apply the particular notation 
$U^{\pm c}_{ab}(\bar s)$ as introduced in (\ref{res-varsigmaA}).

It should not come as a surprise that like the $D$-  also the $B$-function does not satisfies the Schwarz reflection principle  with
\begin{eqnarray}
 B^*_{ab}(s) \neq B_{ab}(s^*)\,,
 \label{B-Schwarz}
\end{eqnarray}
even in the limit with $\epsilon \to 0$. This is readily verified. If expanded to second order in powers of the generalized potential there must be an anomalous contribution with (\ref{B-Schwarz}) that cancels the effect of the anomalous box contribution (\ref{res-Ubox}). This is so since by construction the full reaction amplitude was constructed to satisfy (\ref{def-Schwarz}) at least to second order in a perturbative expansion.

\section{Linear integral equation on real contours}

In the previous two sections we have expressed the $D_{ab}(s)$ and $B_{ab}(s)$ functions in terms of the 
spectral weight $ \hat  \varsigma_{ab} (\bar s)$. In this section we wish to establish a set of linear integral equations
for $ \hat \varsigma_{ab} (\bar s)$ given a generalized potential $U_{ab}(s)$. These equations will serve  
as an alternative formulation of (\ref{res-varsigma}), which is suitable for numerical simulations. After some necessary steps detailed in this section we will 
arrive  at an integral equation of the form
\begin{eqnarray}
&&  \hat \varsigma (s) = -\hat U(s)\,\hat \rho(s) +
\sum_{m,n=1}^3 \int \frac{d \bar s}{\pi }\,
\frac{s-\mu^2}{\bar s-\mu^2}\,\hat \varsigma (\bar s)\,u_m^L(\bar s)\,
\frac{\hat U_{mn}(\bar s)- \hat U_{mn}(s)}{\bar s-s}\,u_n^R(s)\,\hat \rho(s)\,,
\label{res-varsigmahat-final}
\end{eqnarray}
in terms of a set of analytic matrix functions $\hat U(s), \hat U_{mn}(s)$ and $u^L_m(s), u^R_m(s)$. The latter will be expressed in terms of the 
generalized potential $U(s)$. Note that in (\ref{res-varsigmahat-final}) we suppressed the coupled-channel indices. The terms are ordered properly so that the 
coupled-channel structure is correctly implied by standard matrix multiplication rules. 

How to cast the contour integral equation (\ref{res-varsigma}) into an integral equation (\ref{res-varsigmahat-final}) where all integrals are on the real axis strictly with 
$\hat \varsigma_{ac}(s) = 0$ for $s > (m_c + M_c)^2$?
Several steps are required. The first task is to express the functions $\varsigma^\pm_{ab}(s)$ in terms of $ \hat \varsigma_{ab}(s)$ only. 
We begin with the consideration of $\varsigma^-_{ab}(s)$, which we evaluate according to 
the second identity in (\ref{def-varsigma}) with 
\begin{eqnarray}
 \varsigma_{ab}(s) = \sum_d\,\Big[ B^{b-}_{ad}( s)  - \sum_c D^{b-}_{ac}( s)\, U^{b-}_{cd}(s)  \Big]\,\rho_{db}(s)\,\qquad {\rm at}\qquad s \in C_b\,,
 \label{def-varsigma-minus}
\end{eqnarray}
for which we consider the contour limit $\epsilon \to 0$ in Fig. \ref{fig:3}. 
The reaction amplitude $T^{b-}_{cd}(\bar s)$ in (\ref{def-varsigma}) is expressed in terms of the 
$B^{b-}_{ab}(\bar s)$ function as evaluated in (\ref{rewrite-B-second}). Similarly for the required $D^{b-}_{ac}(\bar s)$ function we use 
 (\ref{rewrite-D-second}). Then for  $ \epsilon \to 0$ and $s > \hat \mu^2_b$  we find
\begin{eqnarray}
&& \hat \varsigma_{ab}(s) =\varsigma^-_{ab}( s) =  - \sum_c U_{ac}(s)\,\rho_{cb}(s) +\sum_{c,d} \int_{(m_c+ M_c)^2}^\infty \frac{d \bar s }{\pi} \,\frac{s-\mu^2}{\bar s-\mu^2}\,
\frac{\hat \varsigma_{ac}(\bar s)}{\bar s-s }\,\Big[ U_{cd}( \bar s) -U_{cd}(s)\Big] \,\rho_{db} (s)
\nonumber\\
&& \qquad +\, \sum_{c\geq b,d,e}
\int \frac{d \bar s }{\pi} \,\frac{s-\mu^2}{\bar s-\mu^2}\,
\frac{D_{ae}(\bar s+i\,\epsilon')\,\rho^L_{ec}(\bar s)}{\bar s-s } \,\Big[ U_{cd}(\bar s) -  U_{cd}(s) \Big]\,\rho_{db} (s)
\nonumber\\
&& \qquad +\, \sum_{c<b,d,e}
\int \frac{d \bar s }{\pi} \,\frac{s-\mu^2}{\bar s-\mu^2}\,
\frac{D_{ae}(\bar s+i\,\epsilon')\,\rho^L_{ec}(\bar s)}{\bar s-s } \,\Big[ U^{-c}_{cd}(\bar s) -  U_{cd}(s) \Big]\,\rho_{db} (s)
\nonumber\\
&& \qquad -\, \sum_{c< b,d} \int_{\mu_c^2}^{\hat \mu_c^2} \frac{d \bar s }{\pi} \,\frac{s-\mu^2}{\bar s-\mu^2}\,
\frac{\varsigma^+_{ac}(\bar s)}{\bar s-s +i\,\epsilon'}\,\Delta U_{cd}(\bar s) \,\rho_{db} (s)
\nonumber\\
&& \qquad -\, \sum_{e> c> b,d}
\int_{\mu_c^2}^{\hat \mu_b^2} \frac{d \bar s }{\pi} \,\frac{s-\mu^2}{\bar s-\mu^2}\,
\frac{D_{ae}(\bar s+i\,\epsilon)}{\bar s-s+i\,\epsilon' } \,\rho^L_{ec}(\bar s)\,\Delta U_{cd}(\bar s)  \,\rho_{db} (s) \,,
\label{res-varsigma-hat:A}
\end{eqnarray}
where we again use the particular notations $\Delta U_{ab}(\bar s)$ and  
$U^{\pm c}_{ab}(\bar s)$ as introduced in (\ref{def-Delta-U}, \ref{res-varsigmaA}). We emphasize that $s$ and $\bar s$ in (\ref{res-varsigma-hat:A}) are strictly on the real axis as is implied by the limit $\epsilon \to 0$. It is important to realize that $U_{cb}(s)$  is  evaluated in between the upper and lower anomalous  cut lines, the dashed lines in Fig. \ref{fig:2}. The double cut line of the box contribution is not probed here.

We continue with the more complicated case $\varsigma^+_{ab}(s)$ at $s < \hat \mu_b^2$. This time we start with the first identity in (\ref{def-varsigma}) with
\begin{eqnarray}
 \varsigma_{ab}(s) = \sum_d\,\Big[ B^{b+}_{ad}( s)  - \sum_c D^{b+}_{ac}( s)\, U^{b+}_{cd}(s)  \Big]\,\rho_{db}(s)\,\qquad {\rm at}\qquad s \in C_b\,,
 \label{def-varsigma-plus}
\end{eqnarray}
and again consider the contour limit $\epsilon \to 0$. 
For an evaluation of $\varsigma^+_{ab}(s)$ we will need the $D_{ac}(s)$ and $B_{ab}(s)$ functions evaluated slightly above the contour $C_b$ 
with $s \in C_b$. The latter can not be deduced directly from the results of the previous two sections. This is so because sometimes the functions are required in between 
two right-hand cut lines, for which the results (\ref{rewrite-D-second}, \ref{rewrite-B-second}) can not be applied. Some intermediate steps are required. Consider first the $B^{b+}_{ab}( s)$ term, the first contribution in (\ref{def-varsigma-plus}).
In the limit $\epsilon \to 0$ we can derive:
\begin{eqnarray}
&& B^{b+}_{ad}( s)  = B_{ad}( s+ i\,\epsilon') 
\nonumber\\ && \qquad \quad \;- \,\sum_{c < b}\, 2\,i\,
\varsigma^+_{ac}( s) \,\Big( U_{cd}(s)-\Delta U_{cd}( s)\Big)\,\Theta \big[ s- \mu_b^2 \big]\,\Theta \big[\hat \mu_c^2- s \big]\,,
\label{B-intermediate}
\end{eqnarray}
where we consider $ s$ in (\ref{B-intermediate}) to be strictly real again. For the required kinematics with $s < \hat \mu_b^2$ and $\epsilon \to 0$ it follows
\begin{eqnarray}
&&  B^{b+}_{ab}( s) = \sum_{c< b} \int_{(m_c+ M_c)^2}^\infty \frac{d \bar s }{\pi} \,\frac{s-\mu^2}{\bar s-\mu^2}\,\left( 
\frac{\varsigma^-_{ac}(\bar s)\Theta[ \bar s -\hat \mu_c^2 ]}{\bar s-s -i\,\epsilon'} 
+
\frac{\varsigma^+_{ac}(\bar s)\Theta[\hat \mu_c^2 -\bar s ]}{\bar s-s -i\,\epsilon_{cb}'}\right)\,U_{cb}(\bar s) 
\nonumber\\
&& \qquad \quad \,+\, \sum_{c\geq  b} \int_{(m_c+ M_c)^2}^\infty \frac{d \bar s }{\pi} \,\frac{s-\mu^2}{\bar s-\mu^2}\,
\frac{\hat \varsigma_{ac}(\bar s)}{\bar s-s -i\,\epsilon'}\,U_{cb}(\bar s) 
\nonumber\\
&& \qquad \quad \,+\, \sum_{c\geq b,d}
\int \frac{d \bar s }{\pi} \,\frac{s-\mu^2}{\bar s-\mu^2}\,
\frac{D_{ad}(\bar s+i\,\epsilon')\,\rho^L_{dc}(\bar s)}{\bar s-s -i\,\epsilon'} \,U_{cb}(\bar s) 
\nonumber\\
&& \qquad \quad \,+\, \sum_{c< b,d}
\int \frac{d \bar s }{\pi} \,\frac{s-\mu^2}{\bar s-\mu^2}\,
\frac{D_{ad}(\bar s+i\,\epsilon')\,\rho^L_{dc}(\bar s)}{\bar s-s -i\,\epsilon'} \,U^{-c}_{cb}(\bar s)
\nonumber\\
&& \qquad \quad \,-\, \sum_{c< b} \int_{\mu_c^2}^{\hat \mu_c^2} \frac{d \bar s }{\pi} \,\frac{s-\mu^2}{\bar s-\mu^2}\,
\frac{\varsigma^+_{ac}(\bar s)}{\bar s-s -i\,\epsilon_{cb}'}\,\Delta U_{cb}(\bar s) 
\nonumber\\
&& \qquad \quad \, -\, \sum_{c> b,d}
\int_{\mu_c^2}^{\hat \mu_b^2} \frac{d \bar s }{\pi} \,\frac{s-\mu^2}{\bar s-\mu^2}\,
\frac{D_{ad}(\bar s+i\,\epsilon')\,\rho^L_{dc}(\bar s)}{\bar s-s -i\,\epsilon'} \,\Delta U_{cb}(\bar s) 
\,,
\label{rewrite-B-third}
\end{eqnarray}
where we applied (\ref{B-intermediate}) in combination  with  (\ref{rewrite-B-second}). 
 We point the reader to the different 
prescriptions $s \pm i\,\epsilon'$ in the various terms in (\ref{rewrite-B-third}) with
\begin{eqnarray}
 \epsilon_{cb}' = 
\left\{
\begin{array}{lll}
-\,\epsilon' & \quad {\rm for } &  \mu_b^2 < s < \hat \mu_c^2  \\
+\,\epsilon' & \quad {\rm else}
\end{array}
\right. \;\,.
\label{def-epsilon-cb}
\end{eqnarray}
The change in prescription is caused by the 
terms in the right-hand side of (\ref{B-intermediate}). 

We proceed with the second term in (\ref{def-varsigma-plus}). Here we need to evaluate $D^{b+}_{ac}( s)\,U^{b+}_{cb}(s)$ in the limit $\epsilon \to 0$. Progress is based on the identity 
\begin{eqnarray}
&& D^{b+}_{ac}(s) = D_{ac}( s + i\,\epsilon') 
\nonumber\\
&& \qquad \quad\; +\,
\left\{\begin{array}{ll}
- 2\,i\,\varsigma^+_{ac}( s) \,\Theta \big[ s- \mu_b^2 \big]\,\Theta \big[\hat \mu_c^2- s \big] & \qquad {\rm for} \qquad b > c \\
0 &  \qquad {\rm for} \qquad b \leq c
\end{array} \right. \,,
\label{D-intermediate}
\end{eqnarray}
which again follows in the limit $\epsilon \to 0$.
From (\ref{D-intermediate}) it now follows 
\begin{eqnarray}
&&   \sum_{c\geq b} D^{b+}_{ac}( s)\,U^{b+}_{cb}(s) = \sum_{c\geq b} D_{ac}(s+i\,\epsilon')\,U_{cb}(s)\,,
\nonumber\\
&& \sum_{c < b} D^{b+}_{ac}( s)\,U^{b+}_{cb}(s) =  U_{ab}(s)  
 +\sum_{c < b} \int \frac{d \bar s }{\pi} \,\frac{s-\mu^2}{\bar s-\mu^2}\,
\frac{\hat \varsigma_{ac}(\bar s)}{\bar s-s -i\,\epsilon'}\, U_{cb}(s) 
\nonumber\\
&& \qquad \qquad \qquad \quad \;\;\;\,+\,\sum_{c < b, d}
\int\frac{d \bar s }{\pi} \,\frac{s-\mu^2}{\bar s-\mu^2}\,
\frac{D_{ad}(\bar s+i\,\epsilon')\,\rho^L_{dc}(\bar s)}{\bar s-s-i\,\epsilon' }\,U_{cb}(s)  \,.
\label{rewrite-DU}
\end{eqnarray}
where we assumed  $(m_b+M_b)^2 < s < \hat \mu_b^2$ and $\epsilon \to 0$.
Combining our results (\ref{rewrite-B-third}, \ref{rewrite-DU}) we arrive at the desired expression
\begin{eqnarray}
&& \hat \varsigma_{ab}(s) = \varsigma^+_{ab}( s) =   - \sum_c U_{ac}(s)\,\rho_{cb}(s)+ \sum_{c,d} \int_{(m_c+ M_c)^2}^\infty \frac{d \bar s }{\pi} \,\frac{s-\mu^2}{\bar s-\mu^2}\,
\frac{\hat \varsigma_{ac}(\bar s)}{\bar s-s }\,\Big[ U_{cd}(\bar s) -U_{cd}(s)\Big] \,\rho_{db}(s)
\nonumber\\
&& \qquad \;\,+\, \sum_{e>c,d}
\int \frac{d \bar s }{\pi} \,\frac{s-\mu^2}{\bar s-\mu^2}\,
\frac{D_{ae}(\bar s+i\,\epsilon')\,\rho^L_{ec}(\bar s)}{\bar s-s } \,\Big[ \bar U_{cd}(\bar s) - \bar  U_{cd}(s) \Big]\,\rho_{db}(s)
\nonumber\\
&& \qquad \;\,-\, \sum_{c< b} \int_{\mu_c^2}^{\hat \mu_c^2} \frac{d \bar s }{\pi} \,\frac{s-\mu^2}{\bar s-\mu^2}\,
\frac{ \varsigma^+_{ac}(\bar s)}{\bar s-s -i\,\epsilon'_{cb}}\,\Delta U_{cd}(\bar s)\,\rho_{db}(s)
\nonumber\\
&& \qquad \;\,-\, \sum_{e> c>b,d}
\int_{\mu_c^2}^{\hat \mu_b^2} \frac{d \bar s }{\pi} \,\frac{s-\mu^2}{\bar s-\mu^2}\,
\frac{D_{ae}(\bar s+i\,\epsilon)\,\rho^L_{ec}(\bar s)}{\bar s-s-i\,\epsilon' } \,\Delta U_{cd}(\bar s) \,\rho_{db}(s) \,,
\nonumber\\ \nonumber\\
&& \quad {\rm for }\qquad (m_b+M_b)^2 < s < \hat \mu_b^2  \,, \qquad \bar  U_{ab }(\bar s) = U_{ab}(\bar s) \qquad\qquad \;\, {\rm for} \qquad a > b \,,
\nonumber\\ 
&&  \qquad \qquad \qquad \qquad \qquad \qquad \qquad \qquad    \bar  U_{ab }(\bar s) = U_{ab}(\bar s -i \,\epsilon') \qquad {\rm for} \qquad a < b \,,
 \label{res-varsigma-hat:B}
\end{eqnarray}
where we exploited the corresponding prescription changes in (\ref{rewrite-B-third}) and (\ref{rewrite-DU}) that relate to the first line in 
(\ref{res-varsigma-hat:B}). Again we use that $U_{cd}(s-i\,\epsilon') = U_{cd}(s)$ for $ s > (m_d+M_d)^2$ and $c < d$. 
We point at the formal similarity of (\ref{res-varsigma-hat:B}) with our result  (\ref{res-varsigma-hat:A}) derived previously  
at $s > \hat \mu_b^2$ only. 
With the exception of the prescription in the last two terms the expressions are identical. Moreover, since that prescription is relevant only for $s < \hat \mu_b^2$ 
in those terms we have shown that indeed (\ref{res-varsigma-hat:B}) is valid in the extended domain  $s> (m_b+M_b)^2 $.

We turn to $\varsigma^\circ_{ab}(s)\neq \varsigma^+_{ab}(s)$ as was introduced in (\ref{def-varsigma-circ}) and required in the evaluation of $\varsigma^+_{ab}(\bar s)$. In this case we consider the intermediate result 
\begin{eqnarray}
&& \sum_{c < b} D^{b+}_{ac}( s)\,U^{b+}_{cb}(s) =  U_{ab}(s)  + \underbrace{U^{-a}_{ab}(s) - U^{+a}_{ab}(s)}_{\to 0 \quad {\rm for}  \quad b < a} 
\nonumber\\
&& \qquad \qquad +\,\sum_{c < b} \int \frac{d \bar s }{\pi} \,\frac{s-\mu^2}{\bar s-\mu^2}\,
\left( 
\frac{\varsigma^-_{ac}(\bar s)\Theta[ \bar s -\hat \mu_b^2 ]}{\bar s-s -i\,\epsilon'} 
+ \frac{\varsigma^+_{ac}(\bar s)\Theta[\hat \mu_b^2 -\bar s ]}{\bar s-s -i\,\epsilon_{cb}'}\right) 
 \Big(  U_{cb}(s) + U^{-c}_{cb}(s) - U^{+c}_{cb} \Big)
\nonumber\\
&& \qquad \qquad +\,\sum_{c < b, d}
\int\frac{d \bar s }{\pi} \,\frac{s-\mu^2}{\bar s-\mu^2}\,
\frac{D_{ad}(\bar s+i\,\epsilon')\,\rho^L_{dc}(\bar s)}{\bar s-s-i\,\epsilon' }\,U^{-c}_{cb}(s)\,,
\label{rewrite-DU-second}
\end{eqnarray}
where we now assumed  $\mu_b^2< s < (m_b+M_b)^2 $ and $\epsilon \to 0$. After some algebra we find
\begin{eqnarray}
&& \varsigma^\circ_{ab}(s)=  - \sum_c U_{ac}(s)\,\rho_{cb}(s) +\sum_{c,d} \int_{(m_c+ M_c)^2}^\infty \frac{d \bar s }{\pi} \,\frac{s-\mu^2}{\bar s-\mu^2}\,
\frac{\hat \varsigma_{ac}(\bar s)}{\bar s-s }\,\Big[ U_{cd}( \bar s) - U_{cd}(s) \Big] \,\rho_{db} (s)
\nonumber\\
&& \qquad \;\,\,+\, \sum_{e>c,d}
\int \frac{d \bar s }{\pi} \,\frac{s-\mu^2}{\bar s-\mu^2}\,
\frac{D_{ae}(\bar s+i\,\epsilon')\,\rho^L_{ec}(\bar s)}{\bar s-s } \,\Big[\bar  U_{cd}(\bar s) -  \bar U_{cd}(s) \Big]\, \rho_{db} (s)
\nonumber\\
&& \qquad \;\,\,-\, \sum_{c< b,d} \int_{\mu_c^2}^{\hat \mu_c^2} \frac{d \bar s }{\pi} \,\frac{s-\mu^2}{\bar s-\mu^2}\,
\frac{\varsigma^+_{ac}(\bar s)}{\bar s-s -i\,\epsilon_{cb}'}\,\Delta U_{cd}(\bar s) \, \rho_{db} (s)
\nonumber\\
&& \qquad \;\,\,-\, \sum_{e> c> b,d}
\int_{\mu_c^2}^{\hat \mu_b^2} \frac{d \bar s }{\pi} \,\frac{s-\mu^2}{\bar s-\mu^2}\,
\frac{D_{ae}(\bar s+i\,\epsilon)\,\rho^L_{ec}(\bar s)}{\bar s-s-i\,\epsilon' } \,\Delta U_{cd}(\bar s)  \, \rho_{db} (s)\,,
 \label{res-varsigma-hat:C}
\end{eqnarray}
with $ \mu_b^2 <s < (m_b + M_b)^2 $. The reader is pointed at the formal similarity of the two expressions  (\ref{res-varsigma-hat:B}, \ref{res-varsigma-hat:C}). This is in line with our construction with $\varsigma^\circ_{ab}(s) = \hat \varsigma_{ab}(s)$ on $(m_b+M_b)^2< s< \hat \mu_b^2$.

\begin{figure}[t]
\center{
\includegraphics[keepaspectratio,width=0.95\textwidth]{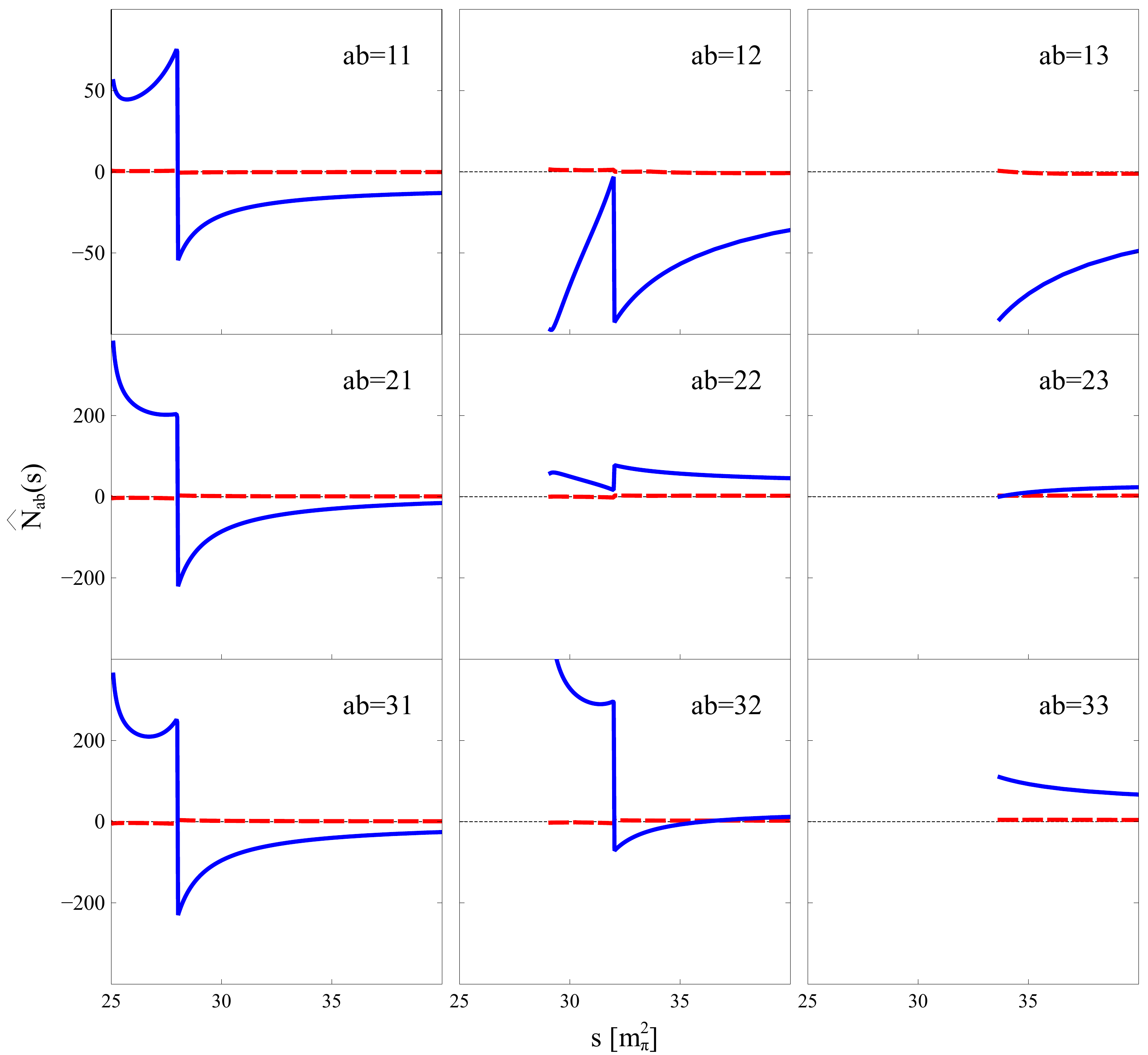} }
\caption{\label{fig:13} The functions $\hat N_{ab}(s)$ of (\ref{res-Nhat}, \ref{def-rhoX}, \ref{res-Nhat-ab}) for $s > \mu_b^2$ or $s > (m_b + M_b)^2 $ in our schematic model (\ref{def-model}). Real and imaginary parts are shown with solid blue and dashed red lines respectively. }
\end{figure}

It remains to rewrite our result (\ref{res-varsigma-hat:B}, \ref{res-varsigma-hat:C}) into a more practical form. This requires three steps. 
First we multiply (\ref{res-varsigma-hat:C}) by the pseudo inverse of the phase-space matrix 
\begin{eqnarray}
&& \hat N_{ab}(s)=  U_{ab}(s) +\sum_{c,d} \int \frac{d \bar s }{\pi} \,\frac{s-\mu^2}{\bar s-\mu^2}\,
\frac{\hat N_{ac}(\bar s)}{\bar s-s }\,\hat \rho_{cd} (\bar s)\,\Big[ U_{db}( \bar s) -U_{db}(s)\Big] 
\nonumber\\
&& \qquad \;\;\;\,-\, \sum_{c>d}
\int \frac{d \bar s }{\pi} \,\frac{s-\mu^2}{\bar s-\mu^2}\,
\frac{D_{ac}(\bar s+i\,\epsilon')\,\rho^L_{cd}(\bar s)}{\bar s-s } \,\Big[ \bar U_{db}(\bar s) -  \bar U_{db}(s) \Big]
\nonumber\\
&& \qquad \;\;\;\,-\, \sum_{b>c} \int \frac{d \bar s }{\pi} \,\frac{s-\mu^2}{\bar s-\mu^2}\,
\frac{ \hat N_{ac}(\bar s)\,\rho^R_{cb} (\bar s) }{\bar s-s -i\,\epsilon_{cb}''}
\nonumber\\
&& \qquad \;\;\;\,\, + \,\sum_{b> c>e}
\int \frac{d \bar s }{\pi} \,\frac{s-\mu^2}{\bar s-\mu^2}\,
\frac{D^{\,c}_{ae}(\bar s)\,\Delta U_{ec}(\bar s) \,\rho^R_{cb}(\bar s) }{\bar s-s -i\,\epsilon_{cb}''} 
\nonumber\\
&& \qquad \;\;\;\, +\, \sum_{c> d > b
} \int \frac{d \bar s }{\pi} \,\frac{s-\mu^2}{\bar s-\mu^2}\,
\frac{D_{ac}(\bar s+i\,\epsilon')\,\rho^{L}_{cd}(\bar s)\,\Delta U_{db}(\bar s)}{\bar s-s-i\,\epsilon' } \,,
 \label{res-Nhat}
\end{eqnarray}
where the result is expressed in terms of the more convenient building blocks
\begin{eqnarray}
&&  \sum_c \hat N_{ac}(s)\, \rho_{cb}(s)= -\Theta \big[ s- (m_b+M_b)^2\big]\, \hat \varsigma_{ab}(s)
-\Theta \big[(m_b+M_b)^2 - s \big]\, \varsigma^\circ_{ab}(s) 
\nonumber\\
&& \qquad\qquad \qquad \quad \;  = \, -\Theta \big[ s- \hat \mu_b^2\big]\, \hat \varsigma_{ab}(s)
-\Theta \big[ \hat \mu_b^2 - s \big]\, \varsigma^\circ_{ab}(s)   \,,
\nonumber\\
&& \hat \rho_{ab}(s) = \rho_{ab}(s) \qquad {\rm  for}\qquad s > (m_a+M_a)^2\,\qquad {\rm  else} \qquad \hat \rho_{ab}(s) = 0\,,
\nonumber\\
&&   \epsilon_{cb}'' = 
\left\{
\begin{array}{lll}
-\,\epsilon' & \quad {\rm for } &  \mu_b^2 < s < \hat \mu_c^2 \quad {\rm or} \quad   s > \hat \mu_b^2  \\
+\,\epsilon' & \quad {\rm else}
\end{array}
\right. \;\,.
\label{def-rhoX}
\end{eqnarray}
The integrals over $\bar s$ in (\ref{res-Nhat}) are on the real axis strictly. The integration domains for the various contributions are implied by 
the region where their integrands are zero as summarized in (\ref{def-rhoL}, \ref{def-rhoX}). Note that a further simplification arises since it is justified to use the replacement $\epsilon_{cb}'' \to -\epsilon'$ in (\ref{res-Nhat}). 
We exemplify the form of the auxiliary functions $\hat N_{ab}(s)$ with our model (\ref{def-model}). In Fig. \ref{fig:13} the complex functions are shown in the domain 
$s> \mu_b^2$ or  $s > (m_b + M_b )^2$ only as needed for the evaluation of the functions $B_{ab}(s)$ in (\ref{rewrite-B-second}, \ref{rewrite-B-third}). It is important to note that the latter do depend on the choice of the return point $\hat \mu_b$. The functions $\hat N_{ab}(s)$ are discontinuous at the return points $s =\hat \mu^2_b$.

In the following we perform a further simplification of (\ref{res-Nhat}). Our target are the terms in the last two lines. 
We begin with the very last term, for which we obtain for $c> b$
\begin{eqnarray}
&& \sum_{c> d > b}\int \frac{d \bar s }{\pi} \,\frac{s-\mu^2}{\bar s-\mu^2}\,
\frac{D_{ac}(\bar s+i\,\epsilon')\,\rho^{L}_{cd}(\bar s)\,\Delta U_{db}(\bar s)}{\bar s-s-i\,\epsilon' } = X^{(a>c>b)}_{ab}(s ) 
\nonumber\\
&& \qquad \qquad \qquad -\,
\sum_{d> c}  \int \frac{d \bar s }{\pi} \,\frac{s-\mu^2}{\bar s-\mu^2}\,
\frac{\varsigma^D_{ad}(\bar s)}{\bar s-s }\,\Big[ X^{(d>c> b)}_{db}(\bar s) - X^{(d>c> b)}_{db}(s) \Big]\,,
\nonumber\\
&& \quad {\rm with}\qquad X^{(a>c>b)}_{ab}(s) =  \int^{\hat \mu^2_b}_{ \mu_c^2} \frac{d \bar s }{\pi} \,\frac{s-\mu^2}{\bar s-\mu^2}\,
\frac{\rho^{L}_{ac}(\bar s)\,\Delta U_{cb}(\bar s)}{\bar s-s - i\,\epsilon' }\,\Theta \Big[ \bar s -(\mu^A_{ac})^2 \Big]\,,
\label{def-XcI}
\end{eqnarray}
where we made the integration domain for the integral representation of $X^{(a>c>b)}_{ab}(s ) $ explicit. This implies that  $X^{(a>c>b)}_{ab}(s ) $ is analytic for $s > \hat \mu_b^2 > (m_b + M_b)^2$ and 
$s < {\rm Max} (\mu_c^2, \mu^A_{ac} )$. Our result (\ref{def-XcI}) follows since the prescription $i\,\epsilon'$ is irrelevant in the $X^{(a>c>b)}_{ab}(\bar s ) $ term. 

An analogous rewrite is  possible for the second last line in (\ref{res-Nhat}). We find
\begin{eqnarray}
&&\sum_{b> c>e}
\int \frac{d \bar s }{\pi} \,\frac{s-\mu^2}{\bar s-\mu^2}\,
\frac{D^{\, c}_{ae}(\bar s)\,\Delta U_{ec}(\bar s) \,\rho^R_{cb}(\bar s) }{\bar s-s +i\,\epsilon'}  = X^{(a<c<b)}_{ab,-}(s ) 
\nonumber\\
&& \qquad \qquad \qquad -\,
\sum_{d< c}  \int \frac{d \bar s }{\pi} \,\frac{s-\mu^2}{\bar s-\mu^2}\,
\frac{\hat \varsigma_{ad}(\bar s)}{\bar s-s }\,\Big[ X^{(d<c< b)}_{db,+}(\bar s) - X^{(d<c<b)}_{db,+}(s) \Big]\,,
\nonumber\\
&& \quad {\rm with}\qquad X^{(a<c<b)}_{ab,\pm}(s) =  \int_{\mu^2_c}^{\hat \mu_a^2} \frac{d \bar s }{\pi} \,\frac{s-\mu^2}{\bar s-\mu^2}\,
\frac{\Delta U_{ac}(\bar s)\,\rho^{R}_{cb}(\bar s)}{\bar s-s \mp i\,\epsilon' }\,\Theta \Big[ \bar s -(\mu^A_{ac})^2 \Big]\,.
\label{def-XcII}
\end{eqnarray}

The results (\ref{def-XcI},\ref{def-XcII}) are useful since they prove that 
the net effect of those terms is a renormalization of the generalized potential of the form
\begin{eqnarray}
 && U^{\rm eff}_{ab,\pm} (s) = U_{ab}(s) + \sum_{a>c>b}\, X^{(a>c>b)}_{ab}(s)
 +  \sum_{a<c<b}\, X^{(a<c<b)}_{ab,\pm}(s)\,, \qquad {\rm and } \qquad  
\nonumber\\ 
&&  \bar U^{\rm eff}_{ab} (s) \;\,  = \bar U_{ab}(s) + \sum_{a>c>b}\, X^{(a>c>b)}_{ab}(s) \,.
 \label{def-UeffI}
\end{eqnarray}
If we use $U^{\rm eff}_{ab,\pm} (s)$ and $\bar U^{\rm eff}_{ab} (s)$   instead of $U^{}_{ab} (s)$ and $\bar U^{}_{ab} (s)$ in the first two lines of (\ref{res-Nhat}) 
the very last term in (\ref{res-Nhat}) can be dropped.
We can combine our results into
\begin{eqnarray}
&& \hat N_{ab}(s)=  U^{\rm eff}_{ab,-}(s) +\sum_{c,d} \int \frac{d \bar s }{\pi} \,\frac{s-\mu^2}{\bar s-\mu^2}\,
\frac{\hat N_{ac}(\bar s)}{\bar s-s }\,\hat \rho_{cd} (\bar s)\,\Big[ U^{\rm eff}_{db,+}( \bar s) -U^{\rm eff}_{db,+}(s)\Big] 
\nonumber\\
&& \qquad \;\;\;\,-\, \sum_{c,d}
\int \frac{d \bar s }{\pi} \,\frac{s-\mu^2}{\bar s-\mu^2}\,
\frac{D_{ac}(\bar s+i\,\epsilon'
)\,\rho^L_{cd}(\bar s)}{\bar s-s } \,\Big[ \bar U^{\rm eff}_{db}(\bar s) - \bar  U^{\rm eff}_{db}(s) \Big]
\nonumber\\
&& \qquad \;\;\;\,-\, \sum_{c} \int \frac{d \bar s }{\pi} \,\frac{s-\mu^2}{\bar s-\mu^2}\,
\frac{ \hat N_{ac}(\bar s)\,\rho^R_{cb} (\bar s) }{\bar s-s +i\,\epsilon'} \,, \qquad \qquad 
 \label{res-Nhat-eff}
\end{eqnarray}
With (\ref{res-Nhat-eff}) we derived a convenient basis for the derivation of a suitable integral equation 
to numerically solve for $\hat \varsigma_{ab}(s)$. After a few more steps we will find 
\begin{eqnarray}
&& \hat N_{ab}(s) = \hat U_{ab}(s) + \sum_{c,d}
 \int \frac{d \bar s}{\pi }\,
\frac{s-\mu^2}{\bar s-\mu^2}\,\hat N_{ac}( \bar s)\,\hat \rho_{cd}(\bar s)\,\hat K_{db}(\bar s,s)\,,
\label{res-Nhat-K}
\end{eqnarray}
with a well behaved integral kernel $\hat K_{db}(\bar s,s)$ and potential term $\hat U_{ab}(s)$. 
It is emphasized that (\ref{res-Nhat-eff}) is valid for $s$ on the real axis strictly, in particular also for 
$s < (m_b+ M_b)^2$.

\begin{figure}[t]
\center{
\includegraphics[keepaspectratio,width=0.95\textwidth]{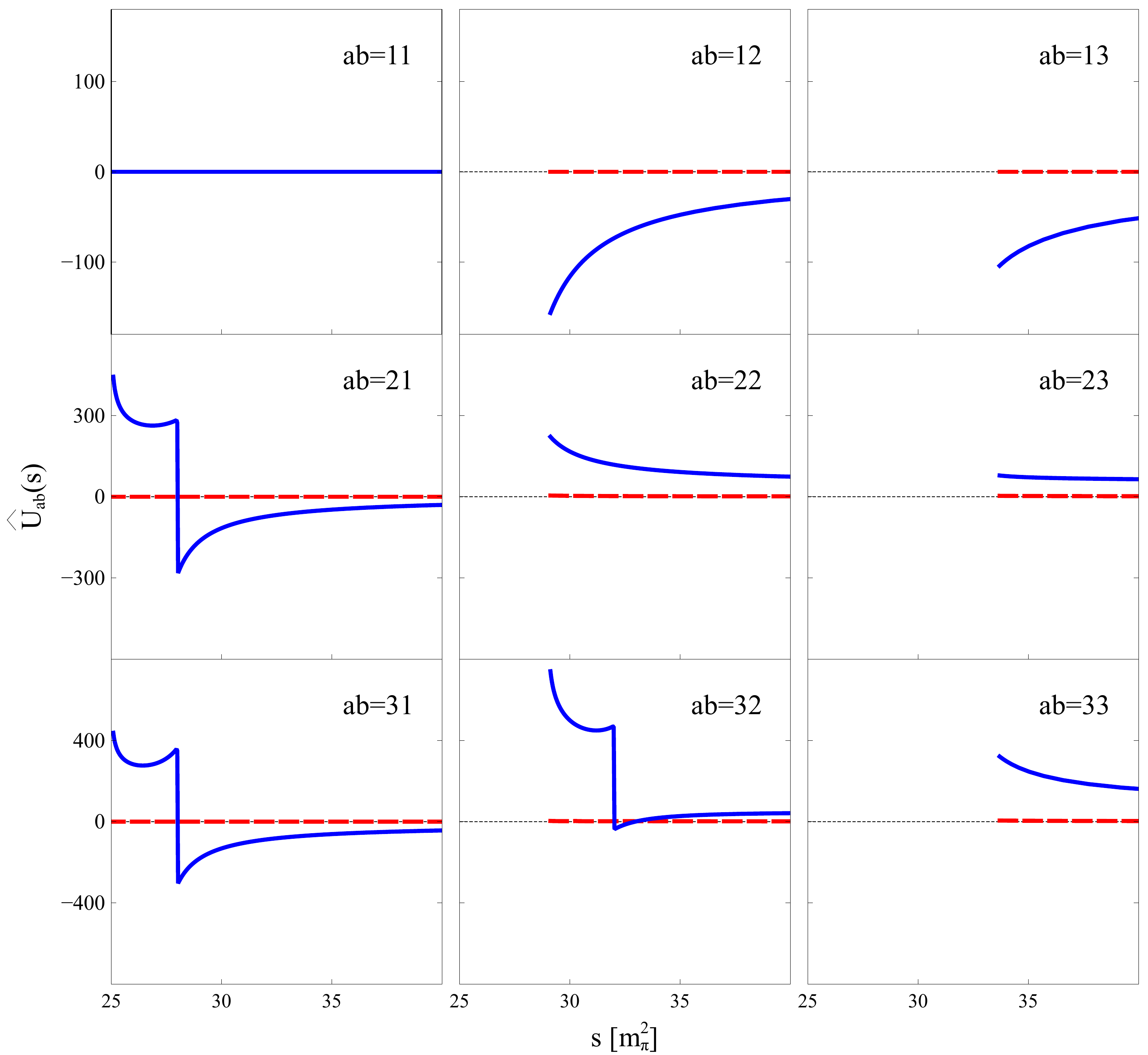} }
\caption{\label{fig:14} The functions $\hat U_{ab}(s)$ of (\ref{res-Nhat-K}, \ref{final-barKinfty}, \ref{res-Uhat}) for $s > \mu_b^2 $ or $s>( m_b + M_b)^2 $ in our schematic model (\ref{def-model}). Real and imaginary parts are shown with solid blue and dashed red lines respectively. }
\end{figure}

The auxiliary functions $\hat U_{ab}(s)$ 
are presented in Fig. \ref{fig:14} as derived from our model (\ref{def-model}). 
The complex functions are shown in the domain $s> \mu_b^2$ or  $s > (m_b + M_b )^2$ only as needed for the evaluation of the functions $\hat N_{ab}(s)$ as shown in Fig. \ref{fig:13}. 
Note that the functions $\hat U_{ab}(s)$ are piecewise continuous in the shown domain with 
the only discontinuous behavior at the return point $s =\hat \mu^2_b$.

The derivation of the anticipated integral equation (\ref{res-Nhat-K}) is organized with the help of the Green's functions $L(x,y)$
and $R(x,y)$. While we already introduced the 'left' Green's function in (\ref{def-Lxy}), the 'right' counter part is readily identified with
\begin{eqnarray}
&& \int dy \bigg[ \delta(x-y) + \frac{1}{\pi}\,\frac{\rho^R(x)}{x-y +i\,\epsilon'} \bigg]\,R (y,z) = \delta(x-z)\,,
\nonumber\\
&& R(x,y) = \delta(x-y) -\frac{1}{\pi}\,\frac{\rho^R(x)}{x-y +i\,\epsilon'} + \sum_{n=1}^3\,\rho^R(x)\,\frac{U_n^R(x)-U_n^R(y)}{x-y}\,\frac{1}{\pi}\,u_n^R(y) \,,
\label{def-Rxy}
\end{eqnarray}
where we emphasize that each of the objects $u_n^R(x)$ and $U_n^R(x)$ has a coupled-channel matrix structure at any $n=1,2,3$. The formal expressions for $u_n^R(x)$ and $U_n^R(x)$ can be extracted 
from our previous expression for $u_n^L(x)$ and $U_n^L(x)$ in (\ref{res-ULa}).  
The close relations amongst the two, left and right,  Green's function is a consequence of the identity
\begin{eqnarray}
 \rho^R_{ab} (x) = \rho^L_{ba}(x)\,,
 \label{rhoR-rhoL}
\end{eqnarray}
which is reflected in (\ref{def-rhoL}, \ref{def-rhoX}).
This implies
\begin{eqnarray}
&& u_n^R(x) = \big[u_n^L(x) \big]^t \,,\qquad  \qquad 
U_n^R(x)  = \big[U_n^L(x) \big]^t \,.
\label{res-URa}
\end{eqnarray}

It remains to recall our previous result (\ref{res-D-from-L}) and apply the right Green's function 
from the right side in (\ref{res-Nhat-eff}).  We multiply the equation with $R(s, x)$ and integrate over $s$. This   
leads to the identification of the potential term $\hat U_{ab}(s)$ and the integral kernel $ \hat K_{ab}(\bar s,s)$ 
of the following form
 \allowdisplaybreaks[1]
\begin{eqnarray}
&& \hat U_{ab}(s ) =   \sum_c \int d \bar s\,
\frac{s-\mu^2}{\bar s-\mu^2}\,\bigg[ U^{\rm eff}_{ac,-}(\bar s)\,R_{cb}(\bar s, s) - \frac{1}{\pi}\, \rho^L_{ac}(\bar s)\, \bar  K_{cb} (\bar s, s)\bigg]\,,
\nonumber\\
&&\bar   K_{ab}(\bar s, s) = \sum_{c,d} \int d x \int d y 
\int L_{ac}(\bar s, x) \, \frac{\bar  U^{\rm eff}_{cd}(x) - \bar U^{\rm eff}_{cd}(y)}{x-y}\,R_{db}(y, s)\,,
\nonumber\\
&& \hat K_{ab}(\bar s, s) = \bar  K_{ab}(\bar s, s)
+ \int d y \,\frac{ U^{\rm eff}_{cd,+}(\bar s) -  U^{\rm eff}_{cd,+}(y)}{\bar s-y}\,R_{db}(y, s)
\nonumber\\
&& \qquad \qquad - \,\int d y \,\frac{\bar  U^{\rm eff}_{cd}(\bar s) - \bar U^{\rm eff}_{cd}(y)}{\bar s-y}\,R_{db}(y, s) \,,
\label{final-barKinfty}
\end{eqnarray}
with which we finally identified the ingredients of (\ref{res-Nhat-K}) in terms of the phase-space matrix $\hat \rho_{ab}(s)$ and the 
generalized potential $U^{\rm eff }_{ab}(s)$ introduced already in (\ref{def-rhoX}, \ref{res-Nhat-eff}).

\begin{figure}[t]
\center{
\includegraphics[keepaspectratio,width=0.95\textwidth]{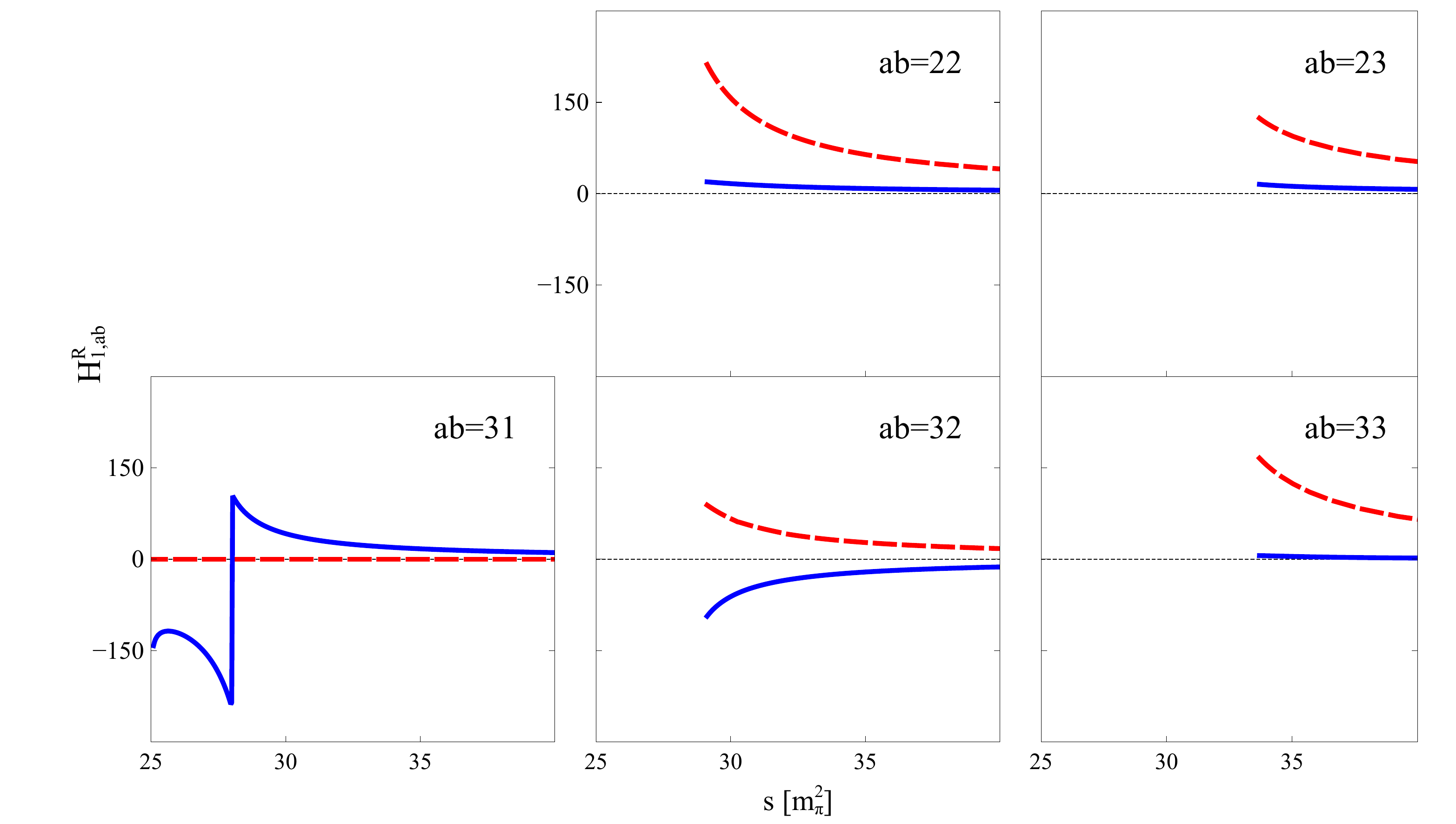} }
\caption{\label{fig:15a} The non-vanishing functions $H^R_{1}(s)= \sum_{m=1}^5 \hat U_{1m}(s) \, u^R_m(s)$ for $s > \mu_b^2$  or $s > (m_b + M_b)^2$ in our schematic model (\ref{def-model}). Real and imaginary parts are shown with solid blue and dashed red lines respectively.  }
\end{figure}

\begin{figure}[t]
\center{
\includegraphics[keepaspectratio,width=0.95\textwidth]{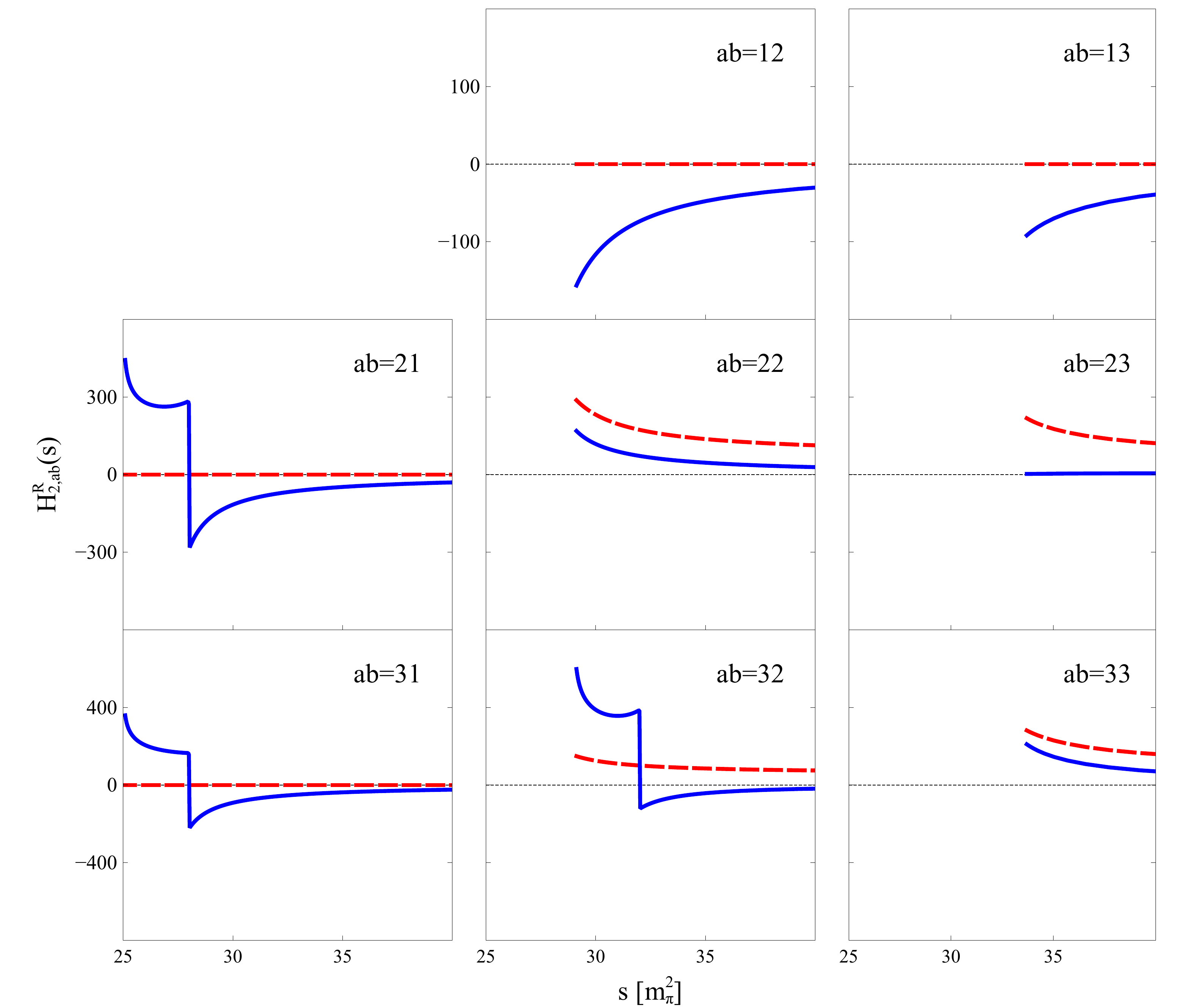} }
\caption{\label{fig:15b} The non-vanishing functions $H^R_{2}(s) =\sum_{m=1}^5 \hat U_{2m}(s) \, u^R_m(s)$ for $s > \mu_b^2$ or $s > (m_b + M_b)^2$ in our schematic model (\ref{def-model}). Real and imaginary parts are shown with solid blue and dashed red lines respectively. }
\end{figure}

\begin{figure}[t]
\center{
\includegraphics[keepaspectratio,width=0.95\textwidth]{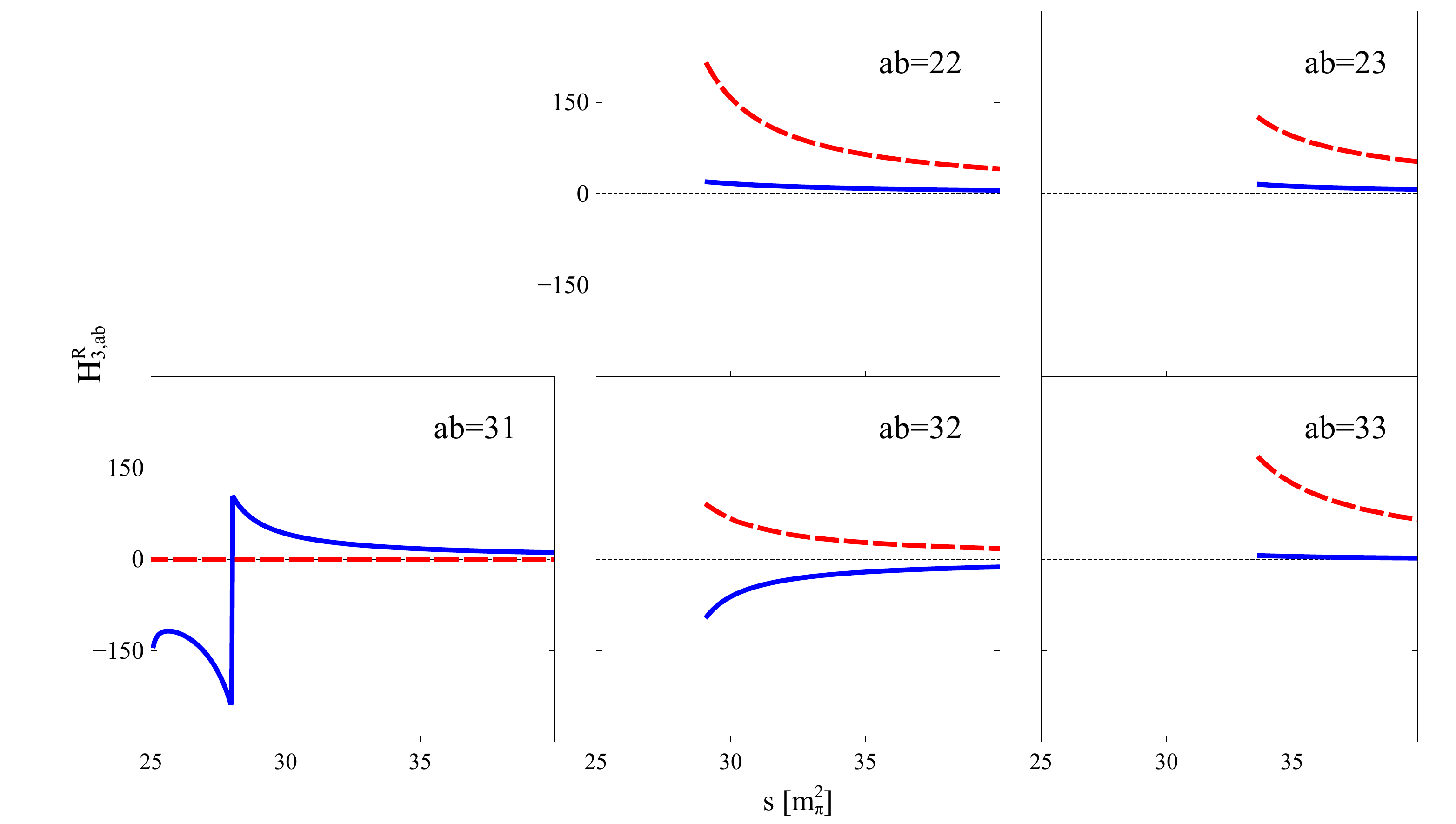} }
\caption{\label{fig:15c} The non-vanishing functions $H^R_{3}(s)=\sum_{m=1}^5 \hat U_{3m}(s) \, u^R_m(s)$ for $s >\mu_b^2$ or $s > (m_b + M_b)^2$ in our schematic model (\ref{def-model}). Real and imaginary parts are shown with solid blue and dashed red lines respectively. }
\end{figure}

It is useful to derive somewhat more explicit expressions. This is readily achieved in terms of 
the  identity
 \allowdisplaybreaks[1]

\begin{eqnarray}
&& \int d z \,\frac{\bar U^{\rm eff}(x)-\bar U^{\rm eff}(z)}{x-z} \,R(z,y) = \sum_{n=1}^5 \frac{\bar U^{\rm eff}_n(x)-\bar U^{\rm eff}_n(y)}{x-y}\,u_n^R(y)\,,
\nonumber\\ 
&& u_4^R(x) =- \sum_{n=1}^3 \,U^R_n(x)\,u_n^R(x) \,,\quad 
 u_5^R(x) = 2\,i\,\rho^R(x)- U^R(x) + \sum_{n=1}^3 \int \frac{dz}{\pi} \,
 \frac{\rho^R(z)\, \Delta U^R_n(z)}{z-x}\,u_n^R(x)  \,,
\nonumber\\ 
&& \bar U^{\rm eff}_{n < 4}(x) = -\bar U^{\rm eff } (x)\,\delta_{n3} + 
\int \frac{dz}{\pi}\, \frac{\Delta \bar U^{\rm eff }(z,x)\,\rho^R(z)}{z-x}\,\delta_{n3}
+
\int \frac{dz}{\pi}\, \frac{\Delta \bar U^{\rm eff }(z,x)\,\rho^R(z)\,U^R_n(z)}{z-x}
\,,
\nonumber\\
&& \bar U^{\rm eff}_4(x) = 
\int \frac{dz}{\pi} \,\frac{\Delta \bar U^{\rm eff }(z,x)\,\rho^R(z)}{z-x}\,,
\qquad \qquad  \qquad  \bar U^{\rm eff}_5(x) = \bar U^{\rm eff}(x)\,,
\label{def-Um}
\end{eqnarray}  
where we use $U^R_{n<4}(x)$ and $u_{n< 4}^R(x)$ as previously introduced in (\ref{def-Rxy}). 
An analogous result holds for the action of the left Green's function $L(x,y)$ on such a structure. This can then be applied to derive the integral kernel
\begin{eqnarray}
&& \bar K(\bar s, s) = \sum_{m,n =1}^5 u_m^L(\bar s)\,\frac{ \bar U_{mn}(\bar s)- \bar U_{mn}(s)}{\bar s-s}\,u_n^R(s) \,, \qquad \qquad 
 u_4^L(x) = -\sum_{n=1}^3\,u_n^L(x) \,U^L_n(x) \,,\qquad 
\nonumber\\ 
&&  u_5^L(x) =
2\,i\,\rho^L(x) - U^L(x) + \sum_{n=1}^3 u_n^L(x) \int \frac{dz}{\pi} \frac{ \Delta U^L_n(z,x)\,\rho^L(z) }{z-x} \,,
\nonumber\\ 
&& \bar U^{m< 4}_{mn}(x) = -\bar U^{\rm eff }_n (x)\,\delta_{m3}+
\int \frac{dz}{\pi}\, \frac{\rho^L(z)\,\Delta \bar U^{\rm eff }_n(z,x)}{z-x}\,\delta_{m3}
+
\int \frac{dz}{\pi}\, \frac{U^L_m(z)\,\,\rho^L(z)\,\Delta \bar U^{\rm eff }_n(z,x)}{z-x}
\,,
\nonumber\\
&& \bar U_{4n}(x) = 
\int \frac{dz}{\pi} \,\frac{\rho^L(z)\,\Delta \bar U^{\rm eff }_n(z,x)}{z-x}\,,
\qquad \qquad  \qquad  \bar U_{5n}(x) = \bar U^{\rm eff}_n(x)\,,
\label{res-Umn}
\end{eqnarray}
in terms of a set of analytic functions $\bar U_{mn}(s)$ and $u^L_m(s), u^R_n(s)$. We note that 
$U_m^L(x), \bar U^{\rm eff}_n(x)$ were already introduced in (\ref{res-ULa}) and (\ref{def-Um}). 
A similar algebra leads to the result
 \allowdisplaybreaks[1]
\begin{eqnarray}
&& \hat K(\bar s, s) = \sum_{m,n =1}^5 u_m^L(\bar s)\,\frac{ \bar U_{mn}(\bar s)- \bar U_{mn}(s)}{\bar s-s}\,u_n^R(s)
+ \sum_{n =1}^5 u_3^L(\bar s)\,\frac{ \hat U_{n}(\bar s)- \hat U_{n}(s)}{\bar s-s}\,u_n^R(s)  
\,, \qquad 
\nonumber\\
&& \hat U^{n< 4}_{n}(x) = -  \hat U_{5} (x)\,\delta_{n3}+
\int \frac{dz}{\pi}\, \frac{\Delta  \hat U_{5}(z,x)\,\rho^R(z)}{z-x}\,\delta_{n3}
+
\int \frac{dz}{\pi}\, \frac{\Delta  \hat U_{5}(z,x)\,\rho^R(z)\,U^R_n(z)}{z-x}
\,,
\nonumber\\
&& \hat U^{}_{4}(x) = 
\int \frac{dz}{\pi} \,\frac{\Delta  \hat U_{5} (z,x)\,\rho^R(z)}{z-x}\,,
\qquad \qquad  \qquad  \hat U_{5}(x) = \bar U^{\rm eff}(x)- U^{\rm eff}_+ (x) \,,
\label{res-Umn}
\end{eqnarray}
where the object $U^{\rm eff}_{+}(x)$ was defined already in (\ref{def-UeffI}). Finally, 
the potential term takes the form
\allowdisplaybreaks[1]
\begin{eqnarray}
&& \hat U(s) =   U^{\rm eff}_{-}(s)
- \int \frac{d \bar s}{\pi }\, \frac{s-\mu^2}{\bar s-\mu^2}\,
\frac{U_{-}^{\rm eff}(\bar s)\,\rho^R(\bar s)}{\bar s-s + i\,\epsilon'}\,
\nonumber\\
&& \qquad \;+\,
\int \frac{d \bar s}{\pi }\, \frac{s-\mu^2}{\bar s-\mu^2}\,\Bigg\{
 \sum_{n=1}^3\,
 U_{-} ^{\rm eff}(\bar s)\,\rho^R(\bar s)\frac{U_n^R(\bar s)-U^R_n(s)}{\bar s-s}\,u_n^R(s)\,
\nonumber\\
&& \qquad \qquad \qquad \quad\;\;\; -\, \sum_{m,n=1}^5 \,\rho^L(\bar s)\,u_m^L(\bar s)\,\frac{\bar U_{mn}(\bar s)-\bar U_{mn}(s)}{\bar s-s}\,u_n^R(s)
\Bigg\}
\,.
\label{res-Uhat}
\end{eqnarray}
Altogether the linear integral equation (\ref{res-Nhat-K}) was cast into the simple form
\begin{eqnarray}
&&  \hat N(s) = \hat U(s) +
\sum_{m,n=1}^5 \int \frac{d \bar s}{\pi }\,
\frac{s-\mu^2}{\bar s-\mu^2}\,\hat N(\bar s)\,\hat \rho(\bar s)\,u_m^L(\bar s)\,
\frac{\hat U_{mn}(\bar s)- \hat U_{mn}(s)}{\bar s-s}\,u_n^R(s) \,,
\nonumber\\
&& \qquad {\rm with} \qquad
\hat U_{mn} (s) = \bar U_{mn}(s) + \delta_{m3}\,\hat U_{n}(s) \,,
\label{res-Nhat-ab}
\end{eqnarray}
where with (\ref{def-rhoX}) we finally arrived at the anticipated result (\ref{res-varsigmahat-final}).

We illustrate the role of the auxiliary matrices $\hat U_{mn}(s)$ as derived from our model (\ref{def-model}). Since there are quite a few functions, we focus on the particular combinations
\begin{eqnarray}
H_n^L (s) =\sum_{m=1}^5  u^L_m(  s)\, \hat U_{mn}( s) \, \qquad {\rm and } \qquad H_n^R(s)= \sum_{m=1}^5 \hat U_{nm}(s) \, u^R_m(s)\,, 
 \label{def-hatUnm-shown}
\end{eqnarray}
as they are the relevant entities in (\ref{res-Nhat-ab}). In Fig. \ref{fig:15a}-\ref{fig:15c} 
we shown for 27 elements of $H_{n,ab }^R (s)$ that they are piece wise continuous for $s > \mu_b^2 $ or $s > (m_b+ M_b)^2$.  The complex functions are shown in the domain $s> \mu_b^2$ or  $s > (m_b + M_b )^2$ only as needed for the evaluation of the functions $\hat N_{ab}(s)$ as shown in Fig. \ref{fig:13}.  Note that $H^R_{n,ab}(s)$ are discontinuous only at the return 
points $s =\hat \mu^2_b$. The analogous property holds for the functions  $H^L_{n,ab}(s)$ at $s> (m_a+M_a)^2$.

We should briefly summarize the general work flow  how to derive the phase-shifts and in-elasticity parameters for a given model interaction. 
Given the driving terms $\hat U(s)$ and $\hat U_{nm}(s)$ together with $u^L_n (s)$ and $u^R_n(s)$ as specified in (\ref{res-ULa}, \ref{res-URa}, \ref{final-barKinfty}-\ref{res-Uhat}) we use the linear set of equations (\ref{res-varsigmahat-final}) to determine the function $\hat \varsigma_{ab}(s)$ for $s > (m_b + M_b)^2$. This is a numerical stable task since all driving terms in (\ref{res-varsigmahat-final}) are sufficiently regular in the needed domain. In the next step we can compute the functions $D_{ab}(s + i\,\epsilon)$ in application of (\ref{res-D-from-L}-\ref{res-ULa}). The functions $B_{ab}(s + i\,\epsilon)$ are evaluated from (\ref{rewrite-B-second}, \ref{rewrite-B-third-old}), 
which, however, requires the knowledge of the functions $\varsigma^+_{ab}(s)$. This goes in two steps. First, given 
the functions $\hat \varsigma_{ab}(s)$ we can compute $\hat N_{ab}(s)$ from (\ref{res-Nhat-ab}) at subthreshold energies $ \mu_b^2 < s < (m_b + M_b)^2$ . Then with
\begin{eqnarray}
 &&  \sum_c \hat N_{ac}(s)\, \rho_{cb}(s)=  -\Theta \big[ s- \hat \mu_b^2\big]\, \hat \varsigma_{ab}(s)
-\Theta \big[ \hat \mu_b^2 - s \big]\, \varsigma^\circ_{ab}(s)   \,,
\end{eqnarray}
the desired object $\varsigma^\circ_{ab}(s)$ is available and we can finalize the derivation of the functions $B_{ab}(s+ i\,\epsilon)$ via (\ref{def-varsigma-circ}) and (\ref{rewrite-B-third-old}). With (\ref{def-XcI}) and (\ref{def-XcII}) it follows
\begin {eqnarray}
&&  B_{ab}(s +  i\,\epsilon) =  -  X^{+L}_{cb}( s) - X^{+R}_{cb}( s)
+
 \sum_{c} \int_{\mu_c^2}^{\hat \mu_c^2} \frac{d \bar s }{\pi} \,\frac{s-\mu^2}{\bar s-\mu^2}\,
\frac{ \hat N_{ac}(\bar s) \,\rho^R_{cb}(\bar s) }{\bar s-s - i\,\epsilon}
\nonumber\\
&& \qquad + \,\sum_c \int_{(m_c+ M_c)^2}^\infty \frac{d \bar s }{\pi} \,\frac{s-\mu^2}{\bar s-\mu^2}\,
\frac{ \hat \varsigma_{ac}(\bar s)}{\bar s-s - i\,\epsilon}\,\Big[ U_{cb}(\bar s )
+ \Delta X^{+L}_{cb} (\bar s, s)+   \Delta X^{+R}_{cb} (\bar s, s) \Big] 
\nonumber\\
&& \qquad +\, \sum_{c,d}
\int \frac{d \bar s }{\pi} \,\frac{s-\mu^2}{\bar s-\mu^2}\,
\frac{D_{ad}(\bar s+i\,\epsilon')\,\rho^L_{dc}(\bar s)}{\bar s-s - i\,\epsilon} \, \Big[ \bar U_{cb}(\bar s )
+ \Delta X^{+L}_{cb} (\bar s, s) \Big]
\nonumber\\
&&\qquad = \,U_{ab}( s ) -  \sum_c\,\hat N_{ac}( s )\,\Big[ \delta_{cb} - 2\,i\,\rho^R _{cb}(s) \Big]
 + \,\sum_c \int_{(m_c+ M_c)^2}^\infty \frac{d \bar s }{\pi} \,\frac{s-\mu^2}{\bar s-\mu^2}\,
\frac{ \hat \varsigma_{ac}(\bar s)\, U_{cb}(s )}{\bar s-s - i\,\epsilon}
\nonumber\\
&& \qquad -\,2\,i\,\sum_c \Delta U_{ac}(\bar s)\rho^R _{cb}(s) +
\sum_{c,d}
\int \frac{d \bar s }{\pi} \,\frac{s-\mu^2}{\bar s-\mu^2}\,
\frac{D_{ad}(\bar s+i\,\epsilon')\,\rho^L_{dc}(\bar s)}{\bar s-s - i\,\epsilon} \, \bar U_{cb}(s )\,,
\nonumber\\ \nonumber\\
&& X^{\pm L}_{ab}( s) = \sum_{a > c > b} 
 \int^{\hat \mu^2_b}_{ \mu_c^2} \frac{d \bar s }{\pi} \,\frac{s-\mu^2}{\bar s-\mu^2}\,
\frac{\rho^{L}_{ac}(\bar s)\,\Delta U_{cb}(\bar s)}{\bar s-s \mp i\,\epsilon'}\,\Theta \Big[ \bar s -(\mu^A_{ac})^2 \Big]\,,
 \nonumber\\
&& X^{\pm R}_{ab}( s) = \sum_{a < c < b} 
\int_{\mu^2_c}^{\hat \mu_a^2} \frac{d \bar s }{\pi} \,\frac{s-\mu^2}{\bar s-\mu^2}\,
\frac{\Delta U_{ac}(\bar s)\,\rho^{R}_{cb}(\bar s)}{\bar s-s \mp i\,\epsilon' }\,\Theta \Big[ \bar s -(\mu^A_{ac})^2 \Big]\,,
\label{rewrite-B-fourth}
\end {eqnarray}
where we introduced $\bar U^{\rm eff}_{ab}(s)$ already in (\ref{def-UeffI}).
In turn the reaction amplitudes $T_{ab}(s + i\,\epsilon ')$ can be reconstructed from (\ref{def-B-function}, \ref{res-T-varsigma}) in terms of the functions $D_{ab}(s+ i\,\epsilon)$, $B_{ab}(s+ i\,\epsilon)$ and $U_{ab}(s+ i\,\epsilon )$. 
The phase shifts and in-elasticity parameters 
are then given by (\ref{def-unitarity}).

\clearpage

\section{Summary and outlook}

In this work we presented a novel framework how to deal with coupled-channel systems in the presence 
of anomalous threshold effects. The framework is formulated  for an arbitrary number of channels and 
is suitable for numerical simulations. We list the main corner stones of our development. 
\begin{itemize}
\item 
Given a generalized potential the coupled-channel reaction amplitudes are defined in terms of 
a set of non-linear integral equation formulated on contours in the complex plane. 
\item 
The analytic structure of the generalized potential in the presence of anomalous threshold effects was clarified. The key observation is the fact that the later must not satisfy the Schwarz reflection principle. 
\item 
We confirm previous studies that a physical approach must consider second order terms in the 
generalized potential. The minimal contributions are identified with the terms that are odd under a Schwarz reflection.
\item 
The non-linear integral equation can be solved numerically by a suitable ansatz with Riemann integrals over real energies only. The specific form of the later was derived for the first time. Explicit expressions for the driving terms were presented for an arbitrary number of channels. 
\item 
A schematic 3-channel model was analyzed in the presence of anomalous thresholds. Along our formal developments all key quantities were illustrated in this model. In particular the reaction amplitdues as well as the phase shifts and in-elasticity parameters were computed and discussed. 
\end{itemize}

Given our framework it is now possible to investigate coupled-channel systems including $J^P=1^-$ and $J^P=\frac{3}{2}^+$ states using realistic interactions. Such systems are notoriously challenging since a plethora of anomalous threshold effects are present. We expect such studies to shed more light on the possible relevance of the hadrogenesis conjecture, which predicts such computations to generate a large part of the hadronic excitation spectrum in QCD.

\vskip1cm
\noindent{\bf Acknowledgments}
\vskip0.5cm
\noindent
M.F.M. Lutz thanks E.E. Kolomeitsev for collaboration at an early stage of the project. 
C.L. Korpa was partially supported by the Hungarian OTKA fund K109462, the GSI theory group and the ExtreMe Matter Institute EMMI at the GSI Helmholtzzentrum f\"ur Schwerionenforschung, Darmstadt, Germany.



\bibliography{bibtex}

\appendix

\end{document}